\documentclass{aa}
\pdfoutput=1
\usepackage{amsmath}
\usepackage{natbib}
\usepackage{units}
\usepackage{color}

\usepackage{graphicx}
\usepackage{txfonts}

\usepackage{xfrac}
\usepackage{siunitx}
\usepackage{booktabs}

\newcommand*\dd[1]{\mathrm{d}#1}

\newcommand{\pd}[2]{\frac{\partial #1}{\partial #2} }

\newcommand{\beq}{\begin{equation}}
\newcommand{\eeq}{\end{equation}}
\newcommand{\bea}{\begin{eqnarray}}
\newcommand{\eea}{\end{eqnarray}}

\newcommand{\Mach}{\mathcal{M}}

\newcommand{\doverdr}[1]{\frac{d #1}{d r}}

\begin{document}

\title{Dynamical friction for supersonic motion\\in a homogeneous gaseous medium}

\author{
Daniel Thun
\and
Rolf Kuiper
\and
Franziska Schmidt
\and
Wilhelm Kley
}

\institute{
Institut f\"ur Astronomie und Astrophysik, Universit\"at T\"ubingen, Auf der
Morgenstelle 10, D-72076 T\"ubingen, Germany\\
\email{\{daniel.thun@, 
         rolf.kuiper@, 
         fr.schmidt@student, 
         wilhelm.kley@\}uni-tuebingen.de}\\
}

\date{Received ;}

\abstract
{
The supersonic motion of gravitating objects through a gaseous ambient medium
constitutes a classical problem in theoretical astrophysics.  Its application
covers a broad range of objects and scales from planetesimals, planets, all kind
of stars, up to galaxies and black holes.  Especially the dynamical friction,
caused by the forming wake behind the object, plays an important role for the
dynamics of the system. To calculate the dynamical friction for a particular
system, standard formulae, based on linear theory are often used.
}
{
It is our goal to check the general validity of these formulae and
provide suitable expressions for the dynamical friction acting on the moving
object, based on the basic physical parameters of the problem, namely the mass,
radius, and velocity of the perturber, the gas mass density, soundspeed, and
adiabatic index of the gaseous medium, and the size of the forming wake or
interaction time, respectively.
}
{
We perform dedicated sequences of high resolution numerical studies of rigid
bodies moving supersonically through a homogeneous ambient medium, and calculate
the total drag acting on the object, which is the sum of gravitational and
hydrodynamical drag.  We study cases without gravity with purely hydrodynamical
drag, as well as gravitating objects. In various numerical experiments, we
determine the drag force acting on the moving body and its dependence on the
basic physical parameters of the problem, as given above.  From the final
equilibrium state of the simulations, we compute for gravitating objects the
dynamical friction by direct numerical integration of the gravitational pull
acting on the embedded object. 
}
{
The numerical experiments confirm the known scaling laws for the dependence of
the dynamical friction on the basic physical parameters as derived in earlier
semi-analytical studies.  As a new important result we find that the shock's
stand-off distance is revealed as the minimum spatial interaction scale of
dynamical friction.  Below this radius, the gas settles into a hydrostatic
state, which -- due to its spherical symmetry -- causes no net gravitational
pull onto the moving body.  Finally,
we derive an analytic estimate for the stand-off distance that can used
conveniently in calculating the dynamical friction force.
}
{}


\maketitle


\section{Introduction}
The motion of gravitating objects through a gaseous ambient medium is a
classical problem in theoretical astrophysics.  Of special interest is the force
acting on the object as it determines typical evolution or equilibration time
scales.  This force is caused by two main mechanisms, the purely hydrodynamic
drag, and by dynamical friction. Even without gravity, the flow of the gas
will be perturbed by the body because it acts as an obstacle and blocks the gas
in front of it.  This causes an exchange of momentum between the embedded body
and the surroundings, leading to a {\it hydrodynamic drag force} on the object.
On earth this gives for example rise to the air resistance of all flying
objects, a drag force that tends to oppose the motion of the body.  In an
astrophysical setting, for a moving massive body its gravitational attraction
onto the surrounding leads to the formation of a wake of higher-than-average
density behind the moving object, and, as a result, to a gravitational pull of
the dense region onto the body.  This additional force is again directed
opposite to the motion of the body and leads to a slow down.  This force is
usually called {\it gravitational drag force} or {\it dynamical friction}
\citep{1943ApJ....97..255C,1999ApJ...513..252O}. Hence, the total drag force
acting on a gravitating body is the sum of the hydrodynamic drag and dynamical
friction.

Its astrophysical application covers a wide range of fields, as pointed out
recently by \citet{2014A&A...561A..84L}.  In the common envelope phase, the
decay time of the secondary companion is driven by dynamical friction
\citep{2008ApJ...672L..41R,2012ApJ...746...74R}, as well as the important
survival rate of planets around evolved stars \citep{2009ApJ...705L..81V}.
Further applications include the settling of massive stars in molecular clouds
\citep{2010ApJ...710..583C}, the drag on a star in the accretion flow around
black holes \citep{2000ApJ...536..663N}, the coalescence of black holes
\citep{2005ApJ...634..921A}, and the migration of planetesimals
\citep{2015ApJ...811...54G}. In the framework of planet evolution,
dynamical friction plays a role in the change of planetary inclination due to
the interaction with the disk \citep{Rein:2012p32161, 2013MNRAS.428..658T}.  The
dynamical interactions of evolved stars that move through the interstellar
medium (ISM) determine the observational properties of the dust in the envelope
\citep{2004ApJ...614..796S, 2011ApJ...734L..26V, Meyer:2014p27280,
Meyer:2014p32100, Meyer:2015p32098}. 

This incomplete list of applications shows that the phenomenon of dynamical
friction is widespread and of general importance.  But still, there is no
satisfying derivation of the force on the object, despite many years of research
\citep[see e.g.][]{2014A&A...561A..84L}.  In this study, we tackle this
important problem of dynamical friction via direct numerical modeling.  We
perform hydrodynamical simulations of a gravitating body moving supersonically
through a homogeneous medium and extract the dynamical friction on the object.
We analyze in detail the scaling of the drag with important physical parameter
of the problem such as Mach number and mass of the object and others, and
compare this to existing formulae for dynamical friction.

The paper is organized as follows: 
In Sect.~\ref{sect:theory}, we summarize the status on formulae for dynamical friction. 
In Sect.~\ref{sect:setup}, we describe the hydrodynamics equations and the physical and numerical setup of the simulations.
In Sect.~\ref{sect:laboratory}, we present comparison simulations to laboratory experiments for non-gravitating objects, moving at supersonic speed through air.
In Sect.~\ref{sect:sims}, we present the simulations of gravitating objects for various different physical parameter and discuss their implication on the dynamical friction. 
In Sect.~\ref{sect:dragforce}, we present a new formula for the drag force.
In Sect.~\ref{sect:summary}, we give a brief summary of the study. 

\section{Analytical estimates of dynamical friction} \label{sect:theory}
A first account of dynamical friction has been given by
\citet{1943ApJ....97..255C} who calculated the dynamical friction of a star
embedded in a stellar cluster.  His results were later adapted to the motion of
stars through the ISM or galaxies through the intergalactic medium.  Here, the
ambient gas is typically treated as a static hydrodynamic continuum which is
transversed by a moving body of mass $M$ and the velocity $V_\infty$.  Using the
impulse approximation, where the moving mass perturbs the surroundings only for
a finite time, \citet{1971ApJ...165....1R} derived a general formula for the
dynamical friction force $F_\mathrm{DF}$ acting on the mass $M$:
\begin{equation} \label{eq:fdrag0} 
F_\mathrm{DF} =  4 \pi \rho_\infty \left(\frac{G M}{V_\infty} \right)^2  \, C_A \,.
\end{equation} 
In Eq.~(\ref{eq:fdrag0}), $\rho_\infty$ denotes the ambient constant density, $G$
is the gravitational constant, and $C_A$ stands for the ``Coulomb logarithm''
\begin{equation} \label{eq:coulomb}
C_A =  \ln \left( s_\mathrm{max} / s_\mathrm{min} \right) \,,
\end{equation}
where $s_\mathrm{min}$ and $s_\mathrm{max}$ correspond to the minimum and
maximum impact parameter of the interaction.  A very similar expression for the
drag on a star moving in a zero temperature medium has been derived by
\citet{1944MNRAS.104..273B} who analyzed the change in momentum by the gas that
is collected in a very thin line behind the object.  Additionally, they derived
relations for the gas accretion rate onto the body.

To apply Eq.~\eqref{eq:fdrag0} to various physical problems, the relevant length
scales $s_\mathrm{min}$ and $s_\mathrm{max}$ need to be known.  Typically, for
$s_\mathrm{max}$ the maximum extension of the medium, or a characteristic
(global) length scale is assumed.  For the minimum length, $s_\mathrm{min}$, the
situation is not so clear and no general accepted recipe exists.  Often, either
the physical radius $R$ of the object is chosen or, in case the body becomes
very small, $s_\mathrm{min} \approx \max{(R,R_\mathrm{A})}$. Here $R_\mathrm{A}$
stands for the accretion radius 
\begin{equation} \label{eq:R_acc}
R_\mathrm{A} = \frac{2 G M}{V_\infty^2} \,.   
\end{equation}
Recently, \citet{2011MNRAS.418.1238C} derived an approximation for
$s_\mathrm{min}$ using ballistic orbit theory, which does not account for the
pressure, however.  Obviously, the question of the correct minimum radius of the
object is still not clarified and remains ambiguous.

While Eq.~\eqref{eq:fdrag0} was derived for a pressureless medium (dust), some
corrections have to be applied in case pressure effects play a role. In the
supersonic case, with $V_\infty > c_\infty$, where $c_\infty$ denotes the
soundspeed of the unperturbed medium, a bow shock forms in front of the object
where (additional) energy can be dissipated.  Using the linearized fluid
equations \citet{1971ApJ...165....1R} and \citet{1980ApJ...240...20R} derive a
correction to the above equation that depends on the Mach number of the object.

In an important work, \citet{1999ApJ...513..252O} extended the above analysis and
studied the time-dependent dynamical friction force acting on a massive object
embedded into a gaseous medium in the sub- and supersonic regime.  Her analysis
resulted in an expression for the coefficient $C_A$, very similar to that of
\citet{1980ApJ...240...20R}, which includes a time-dependent maximum radius,
$s_\mathrm{max} \sim V_\infty t$, where $t $ is the time the object has
interacted with the gas.  Her analysis is based on linear perturbation theory
and useful to study time-dependent phenomena.  The expression for the supersonic
case reads
\begin{equation}\label{eq:ostriker}
C_A  = \ln \left[ \frac{V_\infty t}{s_\mathrm{min}} \, 
\left(\frac{\mathcal{M}^2-1}{\mathcal{M}^2} \right)^{1/2}\right] ~,
\end{equation}
where $\mathcal{M} = V_\infty/c_\infty$ denotes the Mach number of the problem.
Obviously, if we set $s_{max} = V_\infty t$, then Eqs.~\eqref{eq:coulomb} and
\eqref{eq:ostriker} agree with each other in the limit of large $\mathcal{M}$,
because for highly supersonic flows pressure effects become less important.
Here, we are not interested in the time-dependent process but will focus on the
final stationary state.

In addition to these analytic estimates there have been many numerical studies
of moving gravitating objects embedded in a gas.  The first to address this
issue numerically was \citet{1971MNRAS.154..141H} who used a special shock
fitting method to calculate the flow around the body and the accretion rate onto
it.  Later, \citet{1985MNRAS.217..367S} were the first to use modern type fluid
dynamical methods on this problem and performed axisymmetric simulations in
spherical polar coordinates.  In their now classic paper they studied the drag
force and the accretion rate onto the object as a function of the velocity and
found very rough qualitative agreement with Eqs.~\eqref{eq:fdrag0} and
\eqref{eq:coulomb}, when using for $s_\mathrm{min}$ the inner radius of the
computational domain.  Most of the subsequent simulations dealt with open inner
boundary conditions and focused on the mass accretion rate onto the object.
An introductory summary to this so called Bondi-Hoyle-Littleton accretion
process is given in the review article by \citet{2004NewAR..48..843E}.  In this
paper we will not study accretion onto the object and study rigid bodies
only.  The main focus will lie on the accurate computation of the dynamical
friction and -- as a pre-requisite -- on the determination of the minimum
integration radius $s_\mathrm{min}$.

The validity of the formula by Ostriker has been demonstrated by
\citet{2001MNRAS.322...67S} who showed agreement for moderate Mach numbers in
non-linear, isothermal simulations. However, they used a Plummer-type potential
with a smoothing length much larger than $R_\mathrm{A}$.  Later,
\citet{2009ApJ...703.1278K} reanalyzed dynamical friction for extended bodies
through numerical simulations.  However, they did not consider objects with a
more or less rigid surface (like stars) but used again a Plummer-type potential
with a smoothing length, and hence their results may be applicable more to
galaxies moving in the intergalactic medium.  They compare their results to the
formula by \citet{1999ApJ...513..252O} and provide a new fitting formula
involving $R_\mathrm{A}$.  They used an axisymmetric cylindrical coordinate
system, which suffers in terms of limited resolution close to the center.

Starting with \citet{1994ApJ...427..342R} and \citet{1994ApJ...427..351R} there
have been many simulations considering the three-dimensional flow around an
accreting object.  An interesting feature discovered already in these first
studies is the onset of unstable flow when breaking the axial symmetry, an
effect that is most pronounced in planar two-dimensional flows, that are
nonphysical however.  A comprehensive summary of two-dimensional (2D) and
three-dimensional (3D) simulations carried out has been given by
\citet{2005A&A...435..397F}.  They point out that the origin of the instability,
i.e.~under what conditions it is expected and its effect on the drag force, are
still not understood.  In this paper we will avoid the question of instability
and focus on purely axisymmetric flows.


It will be useful to compare the drag formula Eq.~(\ref{eq:fdrag0}) to standard
hydrodynamic drag formula.  From dimensional analyses
\citep{1966hydr.book.....L} one finds for the hydrodynamical drag force acting
on an object with cross section $A$
\begin{equation} \label{eq:drag}
F_\mathrm{hydro} = \frac{1}{2} \, C_D \, \rho_\infty \, V_\infty^2 \, A.
\end{equation}
The dimensionless drag coefficient $C_\mathrm{D}$ contains the details of how
the body and the medium physically interact with each other, e.g.~through
surface effects and the dependence on laminar vs.~turbulent flows.  If we use
the accretion radius to set the interaction cross section $A = \pi R_A^2$,
the ratio of the hydrodynamical drag and the dynamical friction is given by:
\begin{equation}
    \frac{F_\mathrm{hydro}}{F_\mathrm{DF}} = \frac{C_D}{2C_A}
\end{equation}

In this study, we determine the dependence of dynamical friction acting on a
rigid, gravitating body moving supersonically through a homogeneous medium on
the basic physical parameters of the system.  First, we demonstrate that the
dynamical friction scales like Eq.~(\ref{eq:fdrag0}) and we derive an analytic
relation for the up-to-now undetermined minimum length scale $s_\mathrm{min}$ of
the problem, and finally give a convenient expression for the dynamical friction
in general.

\section{Physics and numerics}
\label{sect:setup}
\subsection{Problem setup}
We model gravitating, spherical objects with a given mass $M$ and radius $R$
moving with supersonic speed $V_\infty$ through a homogeneous medium, that is
characterized by its mass density $\rho_\infty$, pressure $p_\infty$, and
temperature $T_\infty$ or soundspeed $c_\infty$, respectively.  As initial
condition, the object is placed instantaneously into the ambient medium and the
evolution of the gaseous system is followed using time-dependent hydrodynamical
simulations.  After the system reaches a steady state we determine the
gravitational pull of the gas onto the object (the dynamical friction) as well
as the hydrodynamic drag. 

\subsection{Equations}
\label{sect:hydro}
We study the motion of the ideal gas by solving the Euler equations 
\begin{align}
    \frac{\partial \rho}{\partial t} + \nabla \cdot (\rho \vec{u}) 
        &= 0 \label{eq:euler1} \\
    \pd{}{t}  \rho \vec{u} + \nabla\cdot(\rho\vec{u}\otimes\vec{u}) + \nabla p 
        &= \rho \, \vec{a}_\mathrm{ext} \label{eq:euler2} \\
    \pd{e}{t} + \nabla \cdot [(e + p) \vec{u}] 
        &= \rho \, \vec{u} \cdot \vec{a}_\mathrm{ext} \label{eq:euler3}. 
\end{align}
Here, $\rho$ denotes the gas mass density, $\vec{u}$ the velocity, $p$ the gas
pressure, and $e = e_\mathrm{kin} + e_\mathrm{th}$ the total (kinetic and
thermal) energy density of the gas.  The acceleration due to external forces
$\vec{a}_\mathrm{ext}$ is given here by the gravitational attraction of the
moving object
\begin{equation}
    \vec{a}_\mathrm{ext}  = - \frac{G M}{r^2} \vec{e}_r
\end{equation}
where $r$ is the distance from the center of the object to the position under
consideration.  We close the equations of hydrodynamics with the ideal gas
equation of state
\begin{equation} \label{eq:caloriceos}
   p = (\gamma-1) e_\mathrm{th} \, .
\end{equation}
Gas pressure, density, and temperature are related via
\begin{equation} \label{eq:thermaleos}
    p = \frac{k_\mathrm{B}}{\mu ~ m_\mathrm{H}} \rho T \, ,
\end{equation}
where  $\mu$ is the mean molecular weight, $k_\mathrm{B}$ the Boltzmann constant
and $m_\mathrm{H}$ the mass of the hydrogen atom.  The soundspeed of the gas is
given by
\begin{equation} \label{eq:soundspeed}
    c_\infty = \sqrt{\gamma \frac{p}{\rho}}.
\end{equation}
In our simulations we use a constant value of the adiabatic exponent $\gamma$.
We study the influence of different $\gamma$-values on the dynamical friction in
dedicated parameter series, see Sect.~\ref{sect:gamma}.

From the given values of the density $\rho_\infty$ and pressure $p_\infty$ of
the initially homogeneous medium, its adiabatic index $\gamma$, and the speed
$V_\infty$ of the moving object, the associated Mach number of the system is
given as
\begin{equation}
{\cal M} = V_\infty/c_\infty.
\end{equation}

\subsection{Numerics}
Simulations are carried out using the open-source code PLUTO
\citep{2007ApJS..170..228M}, version 4.  Part of the simulations were carried
out with a new in-house developed CUDA version for usage on GPUs.  We use a
Runge-Kutta time stepping scheme of second order with a second order
reconstruction of states in space, a van-Leer limiter in the reconstruction
step, and a Harten-Lax-van~Leer solver for the Riemann-Problem.

Equations \eqref{eq:euler1} to \eqref{eq:euler3} are solved in the co-moving
frame of the object, i.e.~the object is at rest in the modeling frame and the
initial velocity of the surrounding homogeneous gas corresponds to the physical
speed of the solid body.  We use a 2D spherical grid $(r,\theta)$ assuming axial
symmetry.  The object is fixed at the origin of the coordinate system and the
gas flows into the negative $z$-direction.  The solid surface of the moving
spherical object is represented as boundary conditions at the radial inner
boundary of the computational domain.  Here, we use reflecting boundary
conditions, corresponding to a non-accreting object.  The computational domain
extends in the radial direction from the radius $R$ of the object up to
$R_\mathrm{domain} = 100 \mbox{ or } 1000~R$.  The symmetry axis of the domain
is aligned with the trajectory of the object.  In the polar direction, the
computational domain extends from \num{0} to $\pi$ using a grid spacing uniform
in angle.  To obtain high resolution at the interesting area around the object,
and to ensure an approximately quadratic grid spacing in the radial and the
polar direction of each grid cell, we use a logarithmic grid spacing in the
radial direction.

At the outer radial boundary $R_\mathrm{domain}$, we set the boundary conditions
according to the gas flow of the surrounding: In the upper hemisphere (for
$\theta \in [0,\pi/2]$), we implemented an inflow boundary condition, i.e.~all
ghost cells are set to the unperturbed values of the surrounding medium.  In the
lower hemisphere (for $\theta \in [\pi/2, \pi]$), we implemented a zero-gradient
boundary condition so that the out-flowing material can leave the computational
domain without reflections.  As already mentioned, we perform simulations in
axial symmetry and therefore at $\theta = 0$ (positive $z$-axis) and at $\theta
= \pi$ (negative $z$-axis) we make use of axisymmetric boundary conditions.

Simulations with a size of the body of $R = 0.1 \mbox{ R}_\odot$ are performed
on a grid consisting of $700 \times 480$ grid cells, simulations with a smaller
size of the body of $R = 0.01 \mbox{ R}_\odot$ are performed on a grid
consisting of $860 \times 400$ grid cells.  We present a corresponding
convergence study in detail in appendix~\ref{sect:res_test}.

\subsection{Assumptions and simplifications}
The numerical experiments are performed to determine the dynamical friction in a
general astrophysical context.  Hence, we do not take into account effects,
which depend on a specific system under investigation, such as radiative heating
and cooling.  The flow is assumed to be inviscid.
Self-gravity of the gas is not taken into account.  Effects of magnetic fields
are not investigated.  The long-evolution problem of a time-dependent slow-down
of the moving body by the acting dynamical friction is not included in this
study, see \citet{2001MNRAS.322...67S} for a discussion on this issue.  The
formula for dynamical friction is derived within the co-moving frame of the
moving body. The object is treated as a rigid body where no accretion through
its surface is allowed. The medium is considered to behave as an ideal gas.  

\subsection{Calculation of the drag force}
The total drag force acting on an object moving through an ambient medium can be
calculated from the momentum balance in the final equilibrium state
\citep{1966hydr.book.....L}.  In our case, the object is moving along the
$z$-direction and we have to consider the $z$-component of Eq.~(\ref{eq:euler2})
which reads in equilibrium
\begin{equation}
     \nabla \cdot (\rho u_z \vec{u}) + \pd{p}{z} - \rho a_z  = 0\, ,
\end{equation}
where $a_z$ denotes the $z$-component of the gravitational acceleration due to
the moving object.  Integrating over the whole volume of the computational
domain we can convert the first two parts into surface integrals and obtain 
\begin{equation} \label{eq:force-balance}
     \int_\mathrm{in} \left[ (\rho u_z \vec{u}) + p \vec{n} \right] \cdot \vec{df}  + \int_\mathrm{Vol} \rho a_z dV
    + \int_\mathrm{out} \left[ (\rho u_z \vec{u}) + p \vec{n} \right] \cdot \vec{df}   
    = 0 \,, 
\end{equation}
where $\vec{df}$ is the surface element and $dV$ the volume element, and the
subscript $in$ and $out$ refer to the inner and outer boundary of the
computational domain.  The first term of Eq.~(\ref{eq:force-balance}) represents
the total hydrodynamical force acting on the surface of the body which is the
sum of a momentum transport, $F^\mathrm{tra}_{in}$, through the body's surface
(e.g. for porous objects or open, accreting bodies) and a pressure force,
$F^\mathrm{prs}_{in}$, acting directly on the body. The second term is the
dynamical friction force on the body, $F_\mathrm{DF}$, which
is obtained by integrating over the whole volume of the domain. The sum of these
two contributions must be balanced by the corresponding momentum transport and
pressure terms at the outer boundary, $F^\mathrm{tra}_\mathrm{out}$ and
$F^\mathrm{prs}_\mathrm{out}$, respectively.  In our case, the first surface integral
in Eq.~(\ref{eq:force-balance}) is taken at the surface of the moving object,
here the radius, $R$, of the spherical body, and the last surface integral at
the outer boundary of the domain, $R_\mathrm{domain}$. In the case of an
impermeable rigid body, $F^\mathrm{tra}_{in} = 0$, and we obtain the final force
balance as
\begin{equation} \label{eq:balance}
     F^\mathrm{prs}_\mathrm{in}  + F_\mathrm{DF}  +
     F^\mathrm{tra}_\mathrm{out}  + F^\mathrm{prs}_\mathrm{out} = 0  \,.
\end{equation}
We calculate all these forces for our simulations and evaluate the
importance of the different contributions. Eq.~(\ref{eq:balance}) shows that
the drag acting on an object can be either evaluated at the inner boundary (plus
gravity) or solely at the outer boundary.

\section{Comparison to laboratory experiments} \label{sect:laboratory}
In computational astrophysics, it is only rarely possible to test the numerical
algorithms against laboratory experiments.  However, in the case of the problem
on hand -- a sphere moving supersonically through a gaseous medium --
experimental data is available for non-gravitating moving bodies.  To check the
validity of our numerical ansatz and prove the accuracy of the code, including
setup, boundary conditions, and the calculation of the drag force, we perform
comparison simulations of a body moving supersonically through a homogeneous
gaseous medium (air), which can be compared to existing data from laboratory
experiments, namely by \citet{1982aafm.book.....V} and
\citet{1967JSpRo...4..822B}.

\begin{figure}[htbp]
\begin{center}
\includegraphics[width=0.49\textwidth]{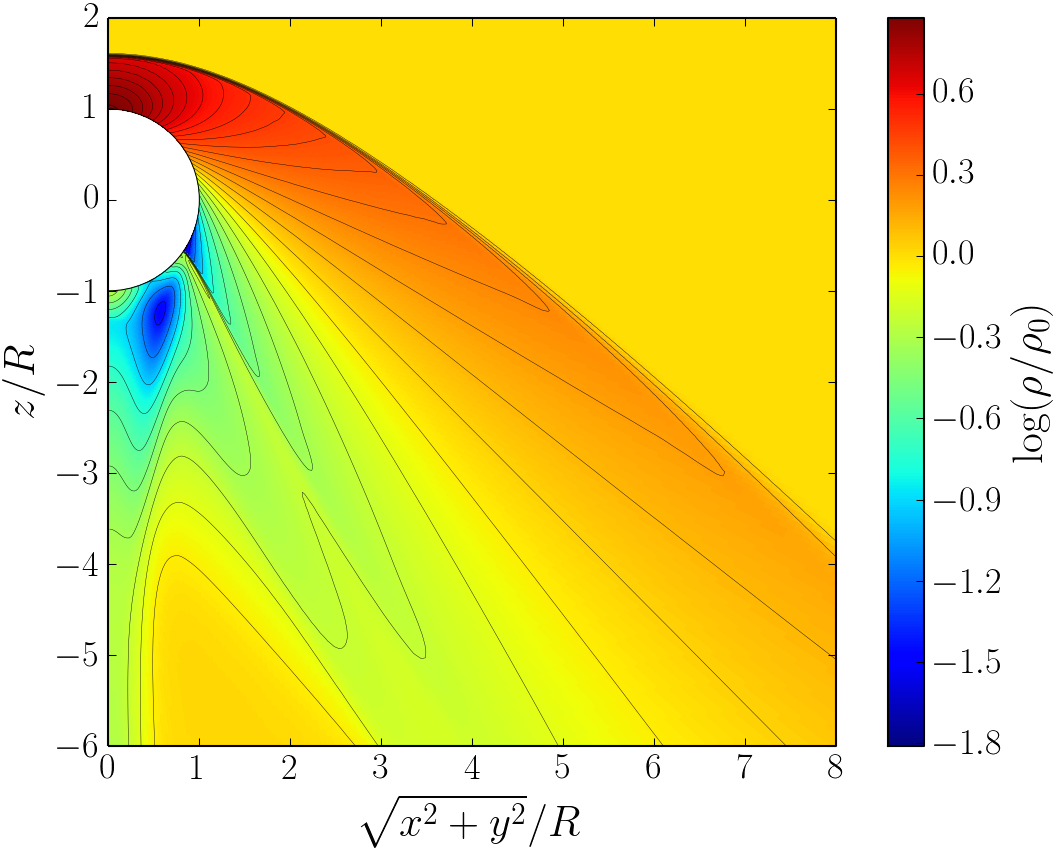} \\
\hspace{-18mm}
\includegraphics[width=0.395\textwidth]{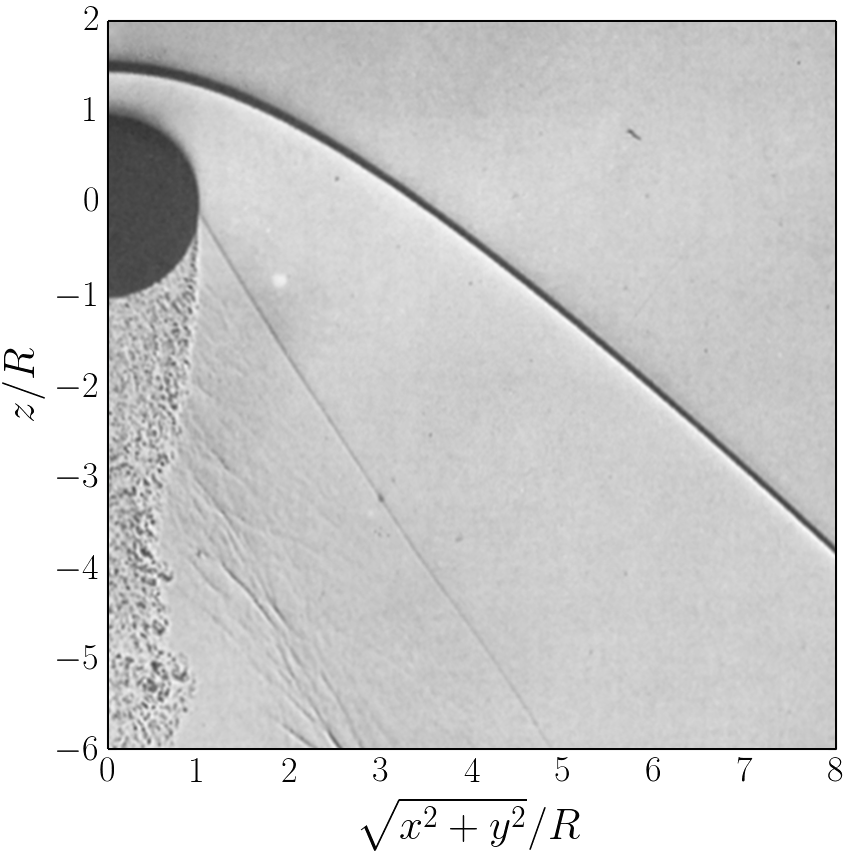} 
\end{center}
\caption{
Visualization of a comparison simulation result of a sphere embedded in air
($\gamma=1.4$) flow at a Mach number $\mathcal{M} = 1.53$.  The homogeneous gas
arrives from the top and flows in the negative $z$-direction.  Shown is the
density distribution for the final equilibrium state in the central part of the
computational domain around the sphere. 
\newline
Upper panel: Numerical results, black lines denote iso-density contours.
\newline
Bottom panel: Laboratory data from \citet{1982aafm.book.....V}, Fig.~266.
}
\label{fig:2d-testflow}
\end{figure}
\subsection{Morphology}
The black-and-white photography in \citet{1982aafm.book.....V}, Fig.~266,
displays the result of a laboratory experiment of a spherical non-gravitating
body moving with supersonic velocity through air.  We present the right half of
the original image in Fig.~\ref{fig:2d-testflow}, bottom panel.  The morphology
of the system is characterized by a standing shock front ahead of the moving
sphere, a clearly visible shock boundary between the unperturbed and the
perturbed gaseous medium, and a low-density region past the moving object.

We model the same experiment numerically within our framework described in the
previous section.  Here, we switch off the gravitational force of the moving
body and use an adiabatic exponent for air of $\gamma = 1.4$.  The body is
moving into the positive $z$-direction; actually, simulations are performed in
the co-moving frame of the body, hence, the gas flow is initialized into the
negative $z$-direction.  The homogeneous gas is initialized with a constant
density $\rho_\infty$.  The Mach number of the flow is set to $\mathcal{M} =
V_\infty/c_\infty = 1.53$.  In the whole computational domain the initial
condition is given by the unperturbed flow.  At the start of the simulation the
rigid sphere is embedded and perturbs the flow.  We run the model until an
equilibrium state has been reached. 

In Fig.~\ref{fig:2d-testflow}, upper panel, we display the morphology of the gas
density around the sphere in the numerical experiment.  Clearly visible is the
bow shock in front of the sphere, where the gas flow changes from supersonic to
subsonic velocities, the shock front dividing the perturbed from the unperturbed
gas, as well as the low-density region behind the moving body.  As expected, the
gas density reaches a maximum directly in front of the sphere.  A visual
comparison with the experimental data from \citet{1982aafm.book.....V},
Fig.~266, as presented in Fig.~\ref{fig:2d-testflow}, shows excellent agreement
between the numerical experiment and the laboratory experiment in terms of their
morphological characteristics.  Furthermore, the magnitude of the density jump
at the shock front from the numerical simulation agrees very well with the
well-known analytical Rankine-Hugoniot jump conditions.  In contrast to the
axisymmetric and inviscid numerical result, the laboratory experiment shows that
the low-density region past the moving object is subject to weak turbulence,
that cannot be captured in the idealized simulation setting.

The following comparison checks for the shock front physics and its dependence
on the Mach number.

\subsection{The stand-off distance of the shock front}
\begin{figure}[htbp]
\begin{center}
\includegraphics[width=0.48\textwidth]{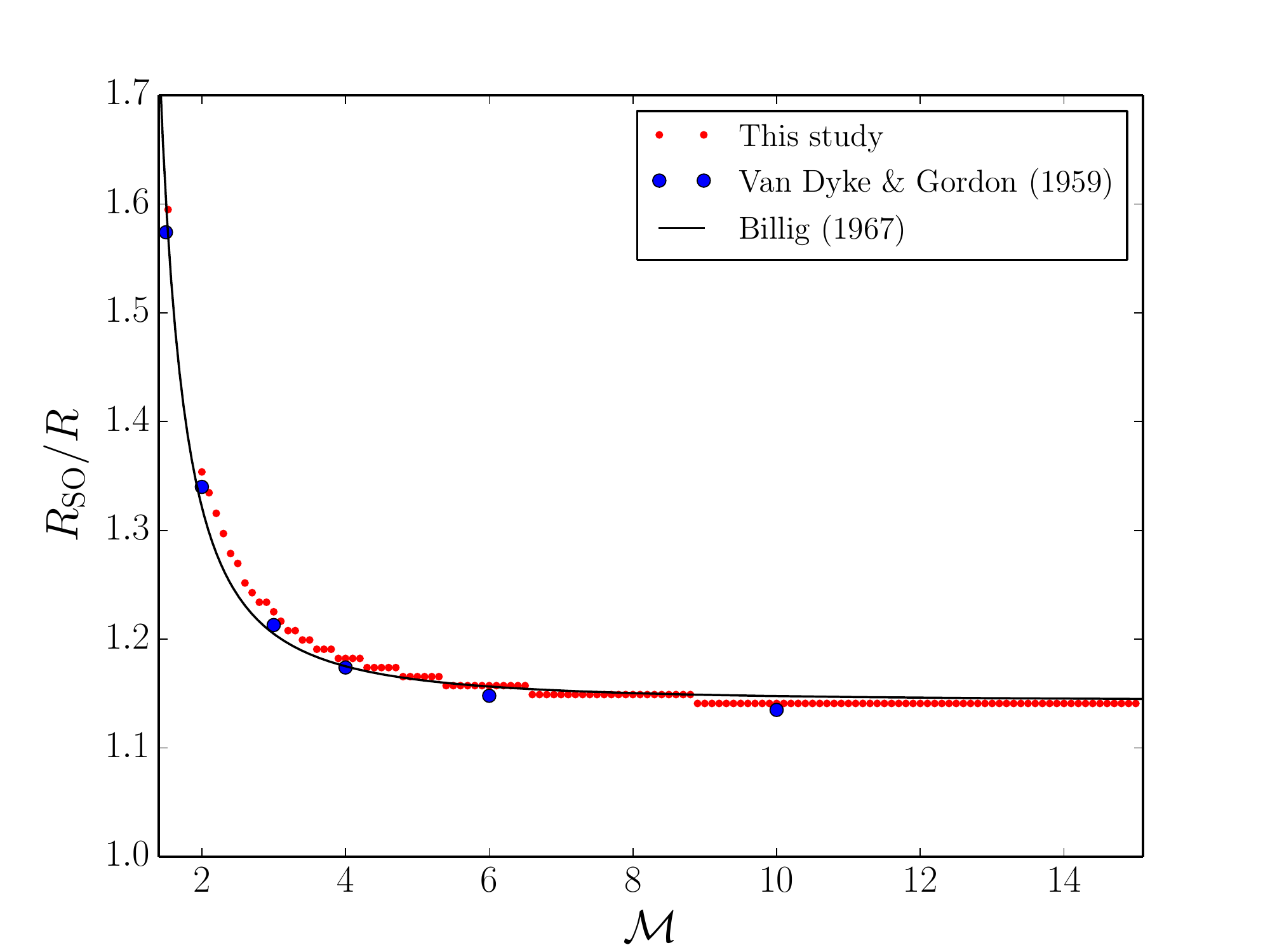}
\end{center}
\caption{
Quantitative comparison of the shock's stand-off distance as function of Mach
number.  The solid black line denotes the relationship derived from laboratory
experiments as given in the main text.  Small red dots represent results from
numerical experiments for the same setup.  Blue dots denote early numerical
results by \citet{VanDykeGordon.1959}.
}
\label{fig:sphere-standoff}
\end{figure}

As we will point out later, the shock's stand-off distance, $R_\mathrm{SO}$,
plays a crucial role in determining the dynamical friction on a gravitating
moving body. Here, we measure $R_\mathrm{SO}$ from the center of the
moving spherical object. To test the dependence of the shock front on the
problem's Mach number, $\mathcal{M}$, we compute a sequence of models for the
same setup as in the previous section but with a variety of different inflow
velocities $V_\infty$.  Fig.~\ref{fig:sphere-standoff} shows the resulting
stand-off distance of the shock, measured from the center of the sphere, as a
function of $\mathcal{M}$.
The $R_\mathrm{SO}$ decreases with increasing $\mathcal{M}$ and approaches
asymptotically in the limit of highly supersonic flow a value of $R_\mathrm{SO}
\rightarrow 1.14$.  This outcome of the numerical experiments can be directly
and quantitatively compared to the laboratory experiments by
\citet{1967JSpRo...4..822B}, using a derived fit to the laboratory data given by
$R_\mathrm{SO}/R \simeq 1 + 0.143 \exp\left(3.24/\mathcal{M}\right)$, where $R$
denotes the geometrical radius of the moving sphere.  The results from the
numerical and laboratory experiments are quantitatively in very good agreement,
as shown in Fig.~\ref{fig:sphere-standoff}.  They are consistent with each other
at both extremes, weak shocks ($\mathcal{M} = 1.53$) as well as strong shocks
($\mathcal{M} \rightarrow \infty$).  In the regime of moderate shocks
($\mathcal{M} \sim 2 \ldots 4$), the numerical experiments show slightly larger
stand-off distances than the laboratory experiments.  Additionally, in the
numerical framework, the position of the shock front is associated with a
certain grid cell, more specifically its center; no further interpolation has
been applied.  Hence, the visible steps in the numerical data visualize the
finite spatial resolution of the numerical experiments. 

\subsection{The hydrodynamical drag force}
\begin{figure}[htbp]
\begin{center}
\includegraphics[width=0.48\textwidth]{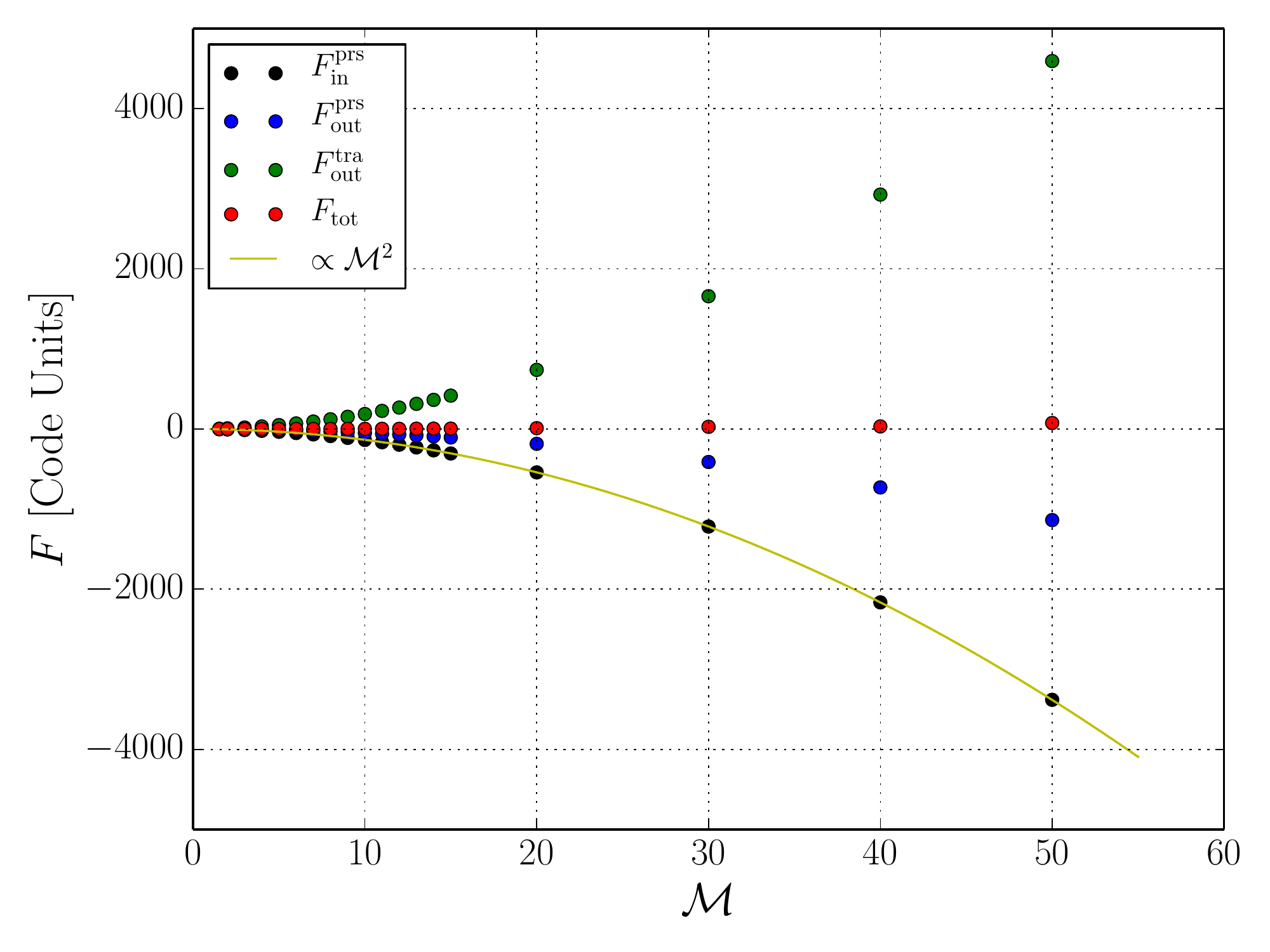}
\end{center}
\caption{
The force contributions in the equilibrium state according to
Eq.~(\ref{eq:balance}) for the non-gravitating, rigid sphere in air, given as a
function of the Mach number.  $F^\mathrm{prs}_\mathrm{in}$ denotes the pressure
force measured at the surface of the sphere, $F^\mathrm{prs}_\mathrm{out}$ the
pressure force measured at the outer boundary of the computational domain, and
$F^\mathrm{trs}_\mathrm{out}$ the momentum transport through the outer boundary.
The sum of these three forces is denoted by $F_\mathrm{tot}$. The curve denotes
a parabola with $F \propto \mathcal{M}^2$.
}
\label{fig:sphere-balance}
\end{figure}
In Fig.~\ref{fig:sphere-balance} we display the contributions of the different
forces acting on the gas and the sphere.  Shown are the pressure force acting on
the body, $F^\mathrm{prs}_\mathrm{in}$, together with the pressure and momentum
flux at the outer boundary.  As to be expected from Eq.~(\ref{eq:balance}) the
sum of all 3 force contributions, $F_\mathrm{tot}$ adds up to zero.  The error
for the total force is slightly increasing for higher Mach numbers but is always
below $2\%$ for all simulations.  In this case the hydrodynamic drag acting on
the body, given by $F^\mathrm{prs}_\mathrm{in}$, is negative which indicates a
force opposing the direction of motion of the body. The standard drag force, as
given in Eq.~(\ref{eq:drag}), depends quadratically on the velocity $V_\infty$
of the object.  Our numerical results clearly confirm this expected quadratic
scaling with $V_\infty$, see added solid line in Fig.~\ref{fig:sphere-balance}.

\section{Simulations of dynamical friction}
\label{sect:sims}
\begin{figure}[t]
\begin{center}
\includegraphics[width=0.49\textwidth]{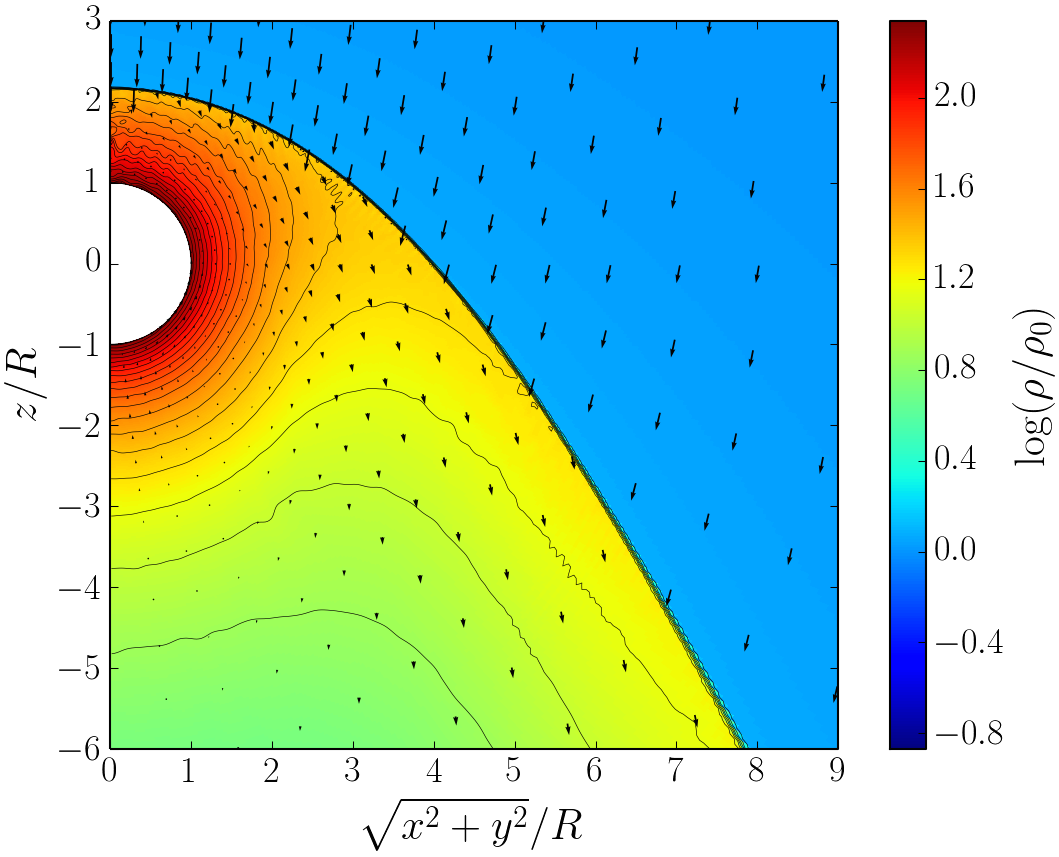}
\end{center}
\caption{
Density and velocity structure of the final quasi-stationary state around a
gravitating object moving with supersonic motion through a gaseous homogeneous
medium.  The body is moving towards the positive $z$-direction. Density is
color-coded and black lines denote iso-density contours. Velocities are given as 
arrows, scaled by the speed.
}
\label{fig:2d-density}
\end{figure}

\renewcommand{\arraystretch}{1.2}
\begin{table*}[tbp]
\centering
\begin{tabular}{l c c c c c c c l}
\#	& $M$ [$\mbox{M}_\mathrm{Jup}$]& $R$ [$\mbox{R}_\odot$] & $\cal{M}$ & $c_\infty$ [$\mbox{ km} \mbox{ s}^{-1}$]& $\gamma$ & $R_\mathrm{domain}$ [$\mbox{R}_\odot$] & Figures \\
\hline\hline
\multicolumn{8}{l}{Fiducial simulation} \\
F	& 60			& ~~0.01 & 4.0 & 58 & 5/3 & 10 & Figs.~\ref{fig:cut_ref} + \ref{fig:stratification}\\
\hline
\multicolumn{8}{l}{Fiducial simulation with larger domain size} \\
C	& 			& ~~0.01 &  &  & & 25 & Fig.~\ref{fig:smax}\\
\hline
\multicolumn{8}{l}{Variation of the mass of the perturber} \\
M1	& 10 \ldots ~~50	& 0.1 & & & &   & Fig.~\ref{fig:eta-dependence-g1.7} + \ref{fig:r_so-r_a} \\
M2	&  \hspace{1ex} 1 \ldots 120	& ~~0.01 & & & &   & Figs.~\ref{fig:rso-dependencies} + \ref{fig:eta-dependence-g1.7} + \ref{fig:scaleM} + \ref{fig:r_so-r_a} \\
\hline
\multicolumn{8}{l}{Variation of the velocity of the perturber} \\
V1	&  			& 0.1 & 3.5 \ldots 20.0 & 49 & 1.2 &   & Fig.~\ref{fig:rso-mach-gamma} + \ref{fig:eta-dependence-g-multi} + \ref{fig:g_factor} \\
V2	&  			& ~~0.01 & 2.2 \ldots 10.9 & 53 & 1.4 &   & Figs.~\ref{fig:rso-mach-gamma} + \ref{fig:eta-dependence-g-multi} + \ref{fig:g_factor} + \ref{fig:r_so-r_a} \\
V3	&  			& 0.1 & 2.0 \ldots ~~9.0 & 58 & 5/3 &   & Figs.~\ref{fig:2d-density} + \ref{fig:eta-dependence-g1.7} + \ref{fig:r_so-r_a} \\
V4	&  			& ~~0.01 & 2.0 \ldots 50.0 & 58 & 5/3 &   &
Figs.~\ref{fig:radial-drag} + \ref{fig:rso-dependencies} +
\ref{fig:eta-dependence-g1.7} + \ref{fig:scaleV} + \ref{fig:rso-mach-gamma} +
\ref{fig:eta-dependence-g-multi} + \ref{fig:g_factor} + \ref{fig:r_so-r_a} +
\ref{fig:df-factor-mach} \\
V5	& 			& ~~0.01 & 2.0 \ldots 20.0 & 63 & 2.0 &  & Figs.~\ref{fig:rso-mach-gamma} + \ref{fig:eta-dependence-g-multi} + \ref{fig:g_factor} + \ref{fig:r_so-r_a} \\
V6	& 			& 0.1 & 2.0 \ldots 10.0 & 77 & 3.0 &   & Figs.~\ref{fig:rso-mach-gamma} + \ref{fig:eta-dependence-g-multi} + \ref{fig:g_factor} \\
V7	& 			& 0.1 & 2.0 \ldots 10.0 & 89  & 4.0 &   & Figs.~\ref{fig:rso-mach-gamma} + \ref{fig:eta-dependence-g-multi} + \ref{fig:g_factor} \\
\hline
\multicolumn{8}{l}{Variation of the soundspeed of the medium} \\
S	& 			& ~~0.01 & 1.6 \ldots 12.6 & 18.3 \ldots 183 & 5/3 &  & Figs.~\ref{fig:eta-dependence-g1.7} + \ref{fig:r_so-r_a} \\
\hline
\multicolumn{8}{l}{Variation of the adiabatic index of the medium} \\
G	& 			& 0.1 & 4.4 & 53 & 1.2 \ldots 6.0 &   & Fig.~\ref{fig:gam-rso} + \ref{fig:g_factor} 
\end{tabular}
\caption{
Overview of series of simulations performed.
Each row represents a series of individual simulations, mostly varying a single basic setup parameter.
The columns denote from left to right
the label of the simulation series,
the mass,
the radius, and
the velocity of the perturber,
the soundspeed of the gaseous medium,
the adiabatic index of the medium, and
the outer radius of the computational domain.
The last column gives the figure numbers associated with the data processing of the simulation series.
The body velocity is given in units of Mach.
If the value of a parameter is not explicitly specified, the value from the fiducial simulation, given in the top row, is used.
}
\label{tab:sims}
\end{table*}
\renewcommand{\arraystretch}{1.0}
Now we turn to astrophysical applications and consider the motion of a
gravitating body through a homogeneous gaseous medium.  For supersonic speeds a
shock front ahead of the moving body is produced very similar to the morphology
found in the laboratory experiments for the non-gravitating bodies.
Fig.~\ref{fig:2d-density} shows the steady state solution for the gas mass
density and the velocity field for one of our simulations.  In front of the
object a bow shock forms where the material is decelerated from supersonic to
subsonic speeds.  After passing the shock front, matter close to the object
settles into a hydrostatic envelope.  The major difference to the
non-gravitating case is that behind the moving object a wake of
higher-than-average density instead of lower density is formed, which is a
direct consequence of the gravitational attraction of the body.  In turn, this
wake of higher density yields a gravitationally pull onto the object, which
slows it down.  This is the phenomenon of dynamical friction.

As shown above in Eq.~(\ref{eq:balance}) the total drag acting on a gravitating
rigid body is the sum of the hydrodynamic drag (pressure force on the body,
$F_\mathrm{in}^\mathrm{prs}$) and the dynamical friction,
$F_\mathrm{DF}$.  In our case, behind the shock front a spherical hydrostatic
shell forms around the object such that the total pressure force on the object
is negligible. Hence, the total drag on the object is given solely by
dynamical friction, and in the following we shall concentrate on this part only,
see Fig.~\ref{fig:grav-balance} below. For very high Mach numbers the separation
of the shock from the object's surface becomes very small such that no
hydrostatic envelope can form, and pressure effects will become important again.

The dynamical friction of a body moving with supersonic speed through a gaseous
homogeneous medium denotes a well defined problem, which involves a manageable
amount of dependencies.  In the course of this section, we first give an
overview of the relevant problem parameters and how they are expected to impact
the amount of dynamical friction (Sect.~\ref{sect:params}).  Afterwards, we
present our realized simulation series, each dedicated to investigate the impact
of a specific parameter, and discuss the simulation results in terms of scaling
laws (Sects.~\ref{sect:scaleM} to \ref{sect:gamma}).  Most importantly, the
simulations' outcome reveal the stand-off distance as the minimum spatial scale
of the forming anisotropic density structure around the moving body.  We analyze
and discuss this new aspect in-depth and derive a semi-analytical relation
between the stand-off distance and the accretion radius from fits to the
numerical data in Sect.~\ref{sect:stand-off}.  Finally, we combine these
findings to derive a convenient semi-analytical expression for the dynamical
friction (Sect.~\ref{sect:dragforce}).

We would like to point out again that we do not assume an a priori validity of
the classic drag formula as stated in Eq.~(\ref{eq:fdrag0}).  Our procedure is
to check the scaling of each parameter in Eq.~(\ref{eq:fdrag0}) individually and
confirm the existence of the logarithmic term. After this we demonstrate that
the minimum distance $s_{min}$ in Eq.~(\ref{eq:coulomb}) is closely related to
the stand-off distance of the shock from the object. We then present a new
formula for calculating $s_{min}$ from the basic physical parameter of the
problem.

\subsection{The relevant problem parameters} \label{sect:params}
First of all, the moving object is characterized by its mass $M$ and velocity
$V_\infty$.  The impact of these parameters on the dynamical friction is
analyzed and discussed in Sects.~\ref{sect:scaleM} and \ref{sect:scaleV}.  For
high enough masses or small enough velocities, the smallest scale of interaction
with the gas is set by the gravity of the moving object rather than its
geometrical radius.  In these cases, the moving body can also be treated as a
point mass, as it is usually done in semi-analytical approaches, e.g.~in the
derivation by \citet{1999ApJ...513..252O}.  Its geometrical radius $R$ becomes
important for the interaction with the surrounding gas in cases of either low
mass or high velocities, respectively.  We investigate the impact of the
geometrical radius in Sect.~\ref{sect:stand-off}.

The gaseous homogeneous medium is characterized by two of the four quantities: 
mass density $\rho_\infty$, pressure $p_\infty$, temperature $T_\infty$, or
soundspeed $c_\infty$. The remaining two are then determined
according to Eqs.~\eqref{eq:thermaleos} and \eqref{eq:soundspeed}.  In the
following, we will choose the gas mass density $\rho_\infty$ and soundspeed
$c_\infty$ as the independent variables; the mass density enters directly the
force term of the gravitational pull of the wake onto the moving body and the
soundspeed of the medium sets the Mach number ${\cal{M}} = V_\infty / c_\infty$
of the shock.

The thermodynamics of the gas is controlled by its caloric and thermal equations
of state Eqs.~\eqref{eq:caloriceos} and \eqref{eq:thermaleos} with the adiabatic
index $\gamma$ as the only free parameter.  Most important for the shock
physics, the adiabatic index controls how the gaseous medium reacts in case of
compression and expansion.

While the object is moving through the gaseous medium, a wake of higher density
will form behind the object, which grows in time.  For a known interaction time
$t$, the extent of the wake is constrained by $s_\mathrm{max} = V_\infty t$.  We
discuss the dependence of the dynamical friction on the maximum extent of the
wake in more detail in Sect.~\ref{sect:smax}.

Summarized, the dynamical friction of a massive body moving supersonically
through a gaseous homogeneous medium depends only on the mass of the body $M$,
the velocity of the body $V_\infty$, its geometrical radius $R$, the mass
density $\rho_\infty$ of the gas, its soundspeed $c_\infty$, the adiabatic index
$\gamma$, and the maximum extent of the wake $s_\mathrm{max}$.

In \citet{1971ApJ...165....1R}, the authors give an expression for the
dynamical friction as given in Eq.~\eqref{eq:fdrag0}.  In this formula, the
dynamical friction depends additionally on the minimum length scale
$s_\mathrm{min}$ of the interaction, a result of the spatial integration limits
of the analytical derivation.  From the analysis of the relevant parameters
above, it follows that the parameter $s_\mathrm{min}$ is actually not a free
parameter, but has to be a function of the parameters given above.  In
Sect.~\ref{sect:stand-off}, we compute the minimum length scale of the
interaction from our numerical solutions, associate this length scale with the
shock's stand-off distance, and derive its dependence on the relevant problem
parameters.

In the following, we present the numerical experiments performed.  The
dependence of the dynamical friction on each of the relevant parameters is
determined in a single or multiple dedicated simulation series.
Table~\ref{tab:sims} gives an overview of the simulation series and their
physical and numerical parameters.  Due to the scale-freedom in the equations of
the problem, the gas mass density can be chosen arbitrarily, we use a value of 
$\rho_\infty = 1.5 10^{-4} \mbox{g cm}^{-3}$ 
in all simulations performed.
In the table, the velocity of the body is giving in units of Mach.  Simulations
with varying soundspeed of the medium (series ``S'') use the same velocity of
the perturber, hence yield different Mach numbers as well.  Simulations with
varying adiabatic index of the medium (series ``G'') use varying initial values
of the gas pressure to keep the soundspeed the same in all simulations of the
series.

\subsection{The gas mass density} \label{sect:scalerho}
The dynamical friction should scale linearly with the density of the environment:
\begin{equation} \label{eq:scalerho}
    F_\mathrm{DF} \propto \rho_\infty
\end{equation}
This scaling behavior is a direct consequence of the fact that the total
dynamical friction is given by the sum of all the gravitational pulls from the
environment onto the body.  Each gravitational pull scales linearly with the
density of the environment.  Additionally, the hydrodynamical equations (see
Sect.~\ref{sect:hydro}) are scale-free in density, and, hence, the flow
morphology is independent on the initial density of the medium.

\subsection{The minimum spatial scale of interaction} \label{sect:stand-off}
The total dynamical friction acting on the moving body is given by the spatial
integral over the gravitational pull of the surrounding gaseous medium.
According to Eqs.~\eqref{eq:fdrag0} and \eqref{eq:coulomb}, it is expected to
scale with the value of a minimum spatial interaction scale $s_\mathrm{min}$
according to the Coulomb logarithm:
\begin{equation}
    F_\mathrm{DF} \propto \ln(s_\mathrm{max} / s_\mathrm{min})
\end{equation}
Below this minimum interaction scale, the gaseous medium is assumed to exert no
net gravitational pull on the body.  What actually sets the minimum interaction
scale is an open discussion in the literature.

\subsubsection{A spatial force analysis}
In our study, the approach of direct numerical experiments allows us to properly
determine the minimum spatial interaction scale in case of gaseous media
quantitatively.  In order to derive the impact of each spatial scale
individually, we first calculate the gravitational drag force acting on the
object from the gas inside a specific shell at radius $r_i$ with a radial shell
thickness of $r_{i+1/2} - r_{i-1/2}$:
\begin{equation} \label{eq:f}
\vec{f}(r_i) = -\int_{S(r_i)}\, \frac{G M}{r^2} ~ \rho(\vec{x}) ~ \vec{\hat{e}}_r\;\dd{V} \, ,
\end{equation}
where the volume of shell $S(r_i)$ is given by
\begin{equation}
S(r_i) =  2 \pi \int_0^{\pi} \int_{r_{i-1/2}}^{r_{i+1/2}} ~ r^2 \sin(\theta)\;\dd{r}\;\dd{\theta} \, .
\end{equation}
The total gravitational force acting on the object is given by the sum over all shells:
\begin{equation}
\label{eq:fgrav-num}
\vec{F}^\mathrm{numerical}_\mathrm{DF} = \sum_i \vec{f} (r_i) \, .
\end{equation}
Even though the above Eq.~\eqref{eq:f} is written in vector form, only its
$z$-component is non-zero due to the axial symmetry of the problem.  The net
gravitational drag will lead to a slow-down of the body, which is moving along
the positive $z$-direction.

\begin{figure}[htbp]
\begin{center}
\includegraphics[width=0.49\textwidth]{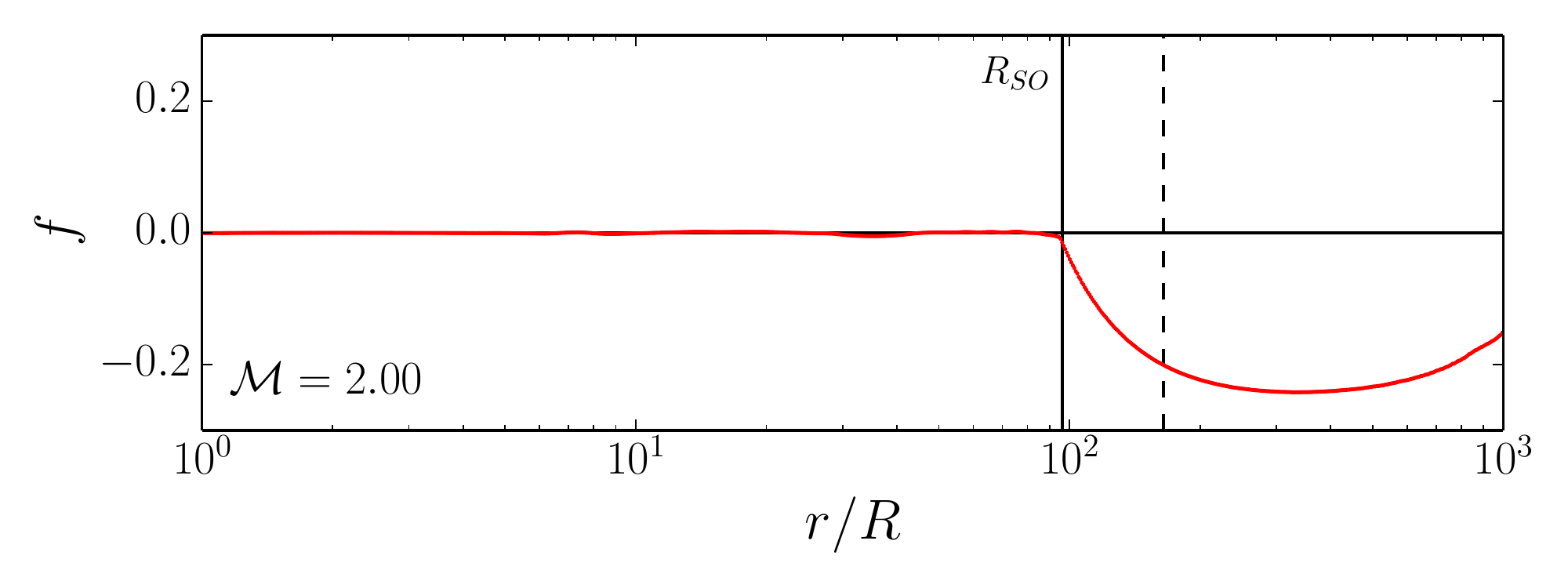}\\
\includegraphics[width=0.49\textwidth]{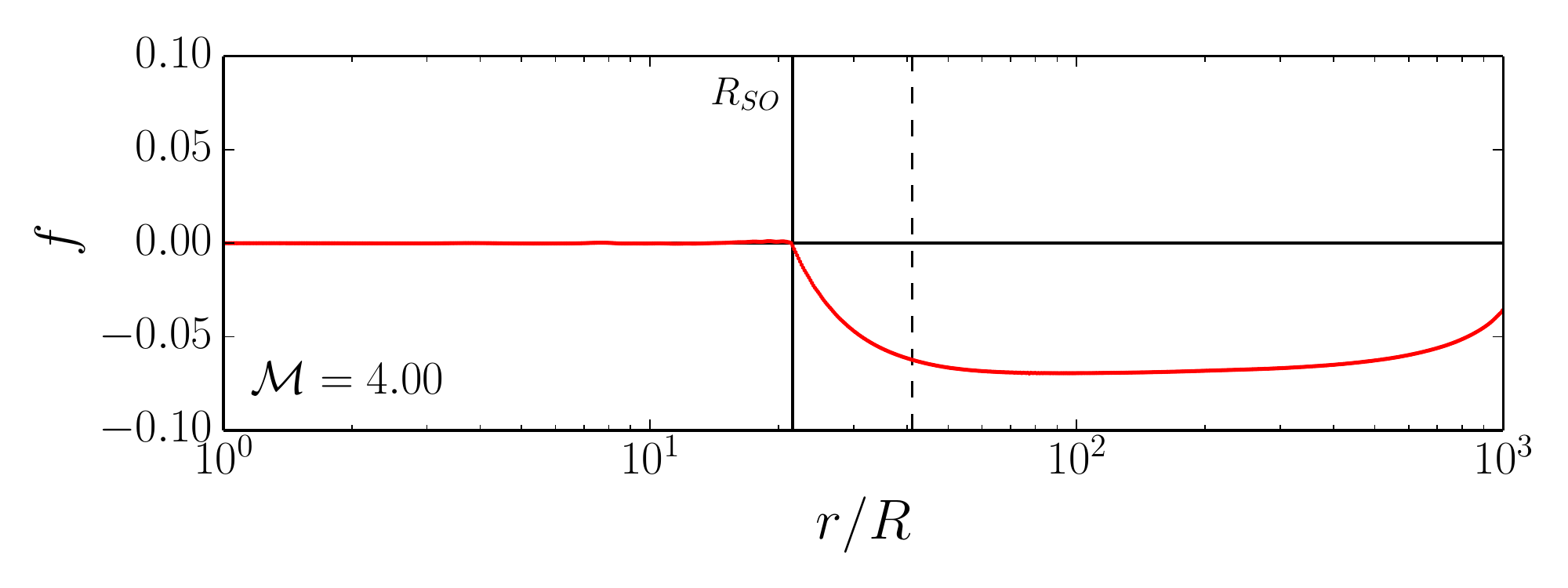}\\
\includegraphics[width=0.49\textwidth]{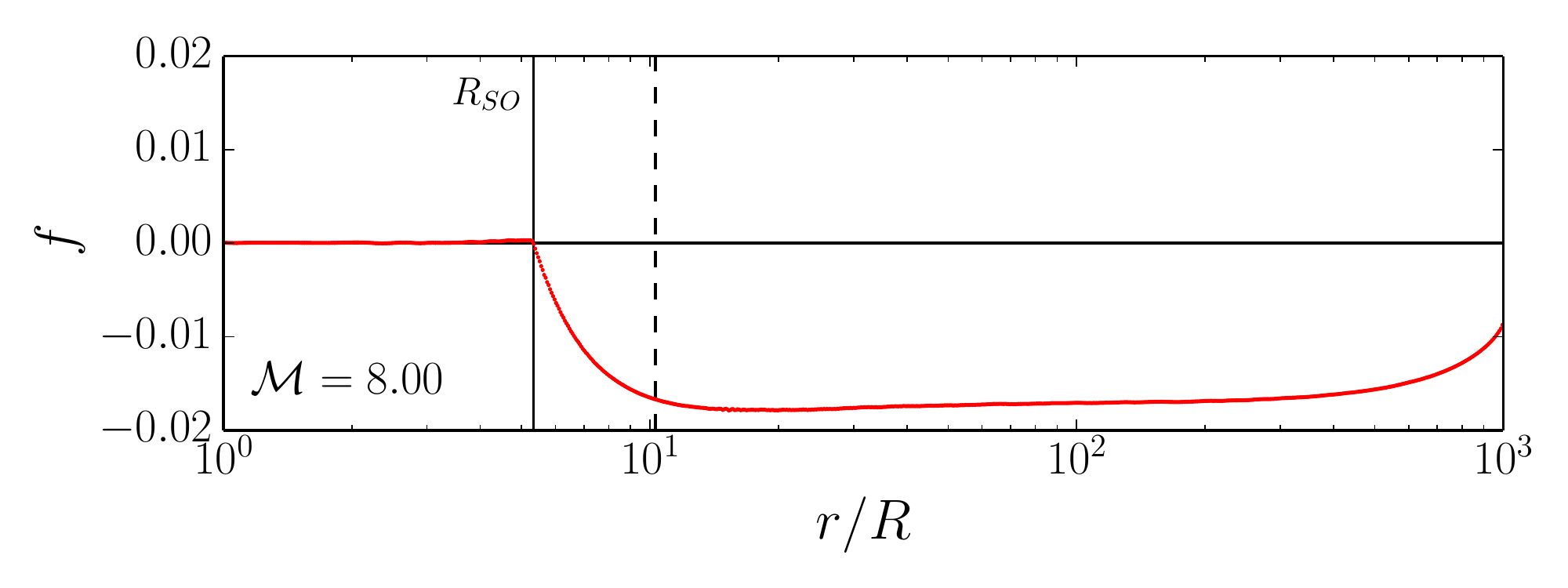}
\end{center}
\caption{
Fractions of the gravitational drag force $f(r_i)$ to the total gravitational
drag force acting on the moving object as a function of distance to the object.
The vertical solid lines mark the stand-off distance $R_\mathrm{SO}$ of the
shock ahead of the object, and the dashed vertical lines denote the accretion
radius $R_\mathrm{A}$ according to Eq.~\eqref{eq:R_acc}.  Results are shown for
simulations with three different Mach numbers $\mathcal{M}$ as labeled in the
bottom left corners. 
}
\label{fig:radial-drag}
\end{figure}

In Fig.~\ref{fig:radial-drag}, we show the fractions of the gravitational force
$f (r_i)$ as a function of radius or distance to the moving body, respectively.
The spatial analysis of the force is shown for three simulations with different
Mach numbers of $\mathcal{M} = 2, 4, \mbox{ and }8$.  Simulation parameters are
given in Table~\ref{tab:sims}, series ``V4''.

The spatial analysis clearly reveals the so-called {\it{stand-off distance}}
$R_\mathrm{SO}$ of the shock front as the searched-for minimum spatial
interaction scale $s_\mathrm{min}$:
\begin{equation}
\label{eq:smin}
s_\mathrm{min}  =  R_\mathrm{SO} \,.
\end{equation}
The associated bow shock forming in front of the body is visible in the density
morphology  depicted in Fig.~\ref{fig:2d-density}.  The shock front
$R_\mathrm{SO}$ denotes the radius, where the flow changes from supersonic to
subsonic velocities.  Moreover, the density morphology around the body is
characterized by the formation of a hydrostatic envelope extending up to
$R_\mathrm{SO}$.  This hydrostatic envelope is very close to spherical symmetry,
as also depicted in Fig.~\ref{fig:cut_ref}.  Hence, although this region marks
the highest gas mass density, and in principle might have the strongest
gravitational impact due to its closeness, the net gravitational pull within
this hydrostatic envelope ($r < R_\mathrm{SO}$) turns out to be negligible.  If
at all, the stronger compression in the forward direction of the trajectory of
the moving body yields a slight deviation from spherical symmetry, which
actually results into a small gravitational acceleration instead of a drag.  But
this effect remains negligible compared to the total dynamical friction and is
only marginally visible in Fig.~\ref{fig:radial-drag} by eye for the case of
highest Mach number; here, the fraction of the dynamical friction force close to
but still below the stand-off distance has a positive sign, indicating positive
acceleration of the body.

\begin{figure}[htbp]
\begin{center}
\includegraphics[width=0.48\textwidth]{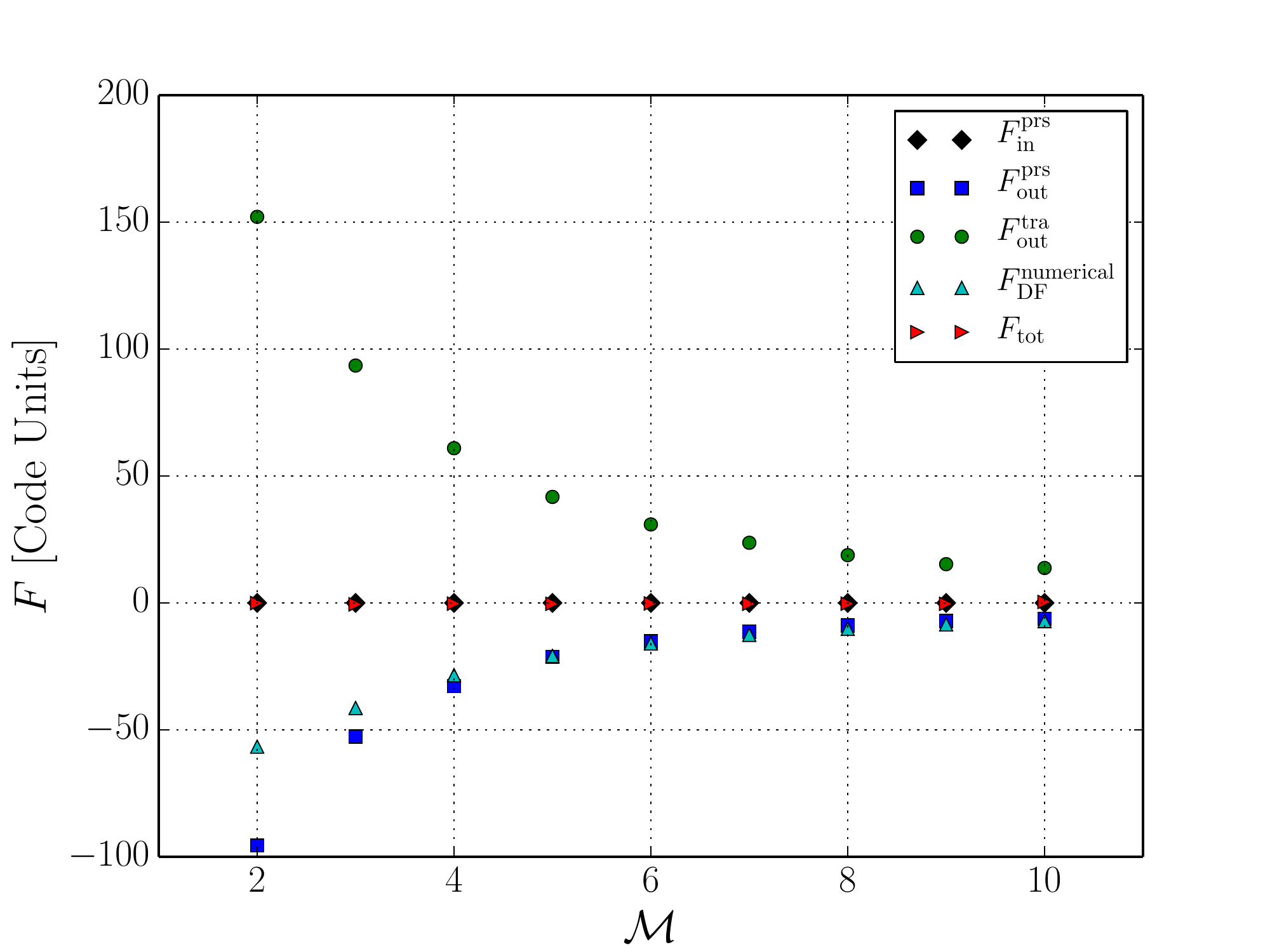}
\end{center}
\caption{
The force contributions for the equilibrium state according to
Eq.~(\ref{eq:balance}) for the {\em gravitating}, rigid sphere, given as a
function of the Mach number.  $F^\mathrm{prs}_\mathrm{in}$ denotes the pressure force
measured at the surface of the sphere, $F_\mathrm{DF}^\mathrm{numerical}$ denotes the total
dynamical friction calculated according to Eq.~(\ref{eq:fgrav-num}),
$F^\mathrm{prs}_\mathrm{out}$ the pressure force measured at the outer boundary of the
computational domain, and $F^\mathrm{trs}_\mathrm{out}$ the momentum transport through
the outer boundary.  The sum of these 4 forces is denoted by $F_\mathrm{tot}$. 
}
\label{fig:grav-balance}
\end{figure}

Clearly, the shells inside the sphere around the object with radius
$R_\mathrm{SO}$ do not contribute to the total dynamical friction force, because
around the object a spherically symmetrical hydrostatic envelope forms.  This is
confirmed in Fig.~\ref{fig:grav-balance} where we plot the individual
contributions to the total drag force on the spherical body, as a function of
Mach number. As shown, the contribution of the pressure force,
$F_{in}^\mathrm{prs}$, acting on the object is negligible. From this we can
conclude that {\em i)} to calculate the drag acting on the object it is
sufficient to consider the dynamical friction alone, and that {\em ii)} the
relevant quantity for the minimum interaction scale $s_\mathrm{min}$ is given by
$R_\mathrm{SO}$.  Additional simulations with fixed Mach number $\mathcal{M}$
but different object masses $M$ give the same result for the spatial analysis of
the force. 

\subsubsection{A convenient expression for the stand-off distance}
It is the aim of this study to derive a general expression for the dynamical
friction, that allows to compute the acting gravitational drag from the relevant
problem parameters without the need of direct numerical simulations.  As
revealed in the previous section, such an expression for the dynamical friction
will include a scaling with
\begin{equation}
F_\mathrm{DF} \propto \ln(s_\mathrm{max} / R_\mathrm{SO})~.
\end{equation}
In the next step of our investigation, we derive an expression for the stand-off
distance, which is based on the basic problem parameters only.  Therefore, we
make use of the apparent relationship of the stand-off distance $R_\mathrm{SO}$
to the accretion radius $R_\mathrm{A}$, see Eq.~\eqref{eq:R_acc} for the
definition of the accretion radius.

\begin{figure}[htbp]
\begin{center}
\includegraphics[width=0.48\textwidth]{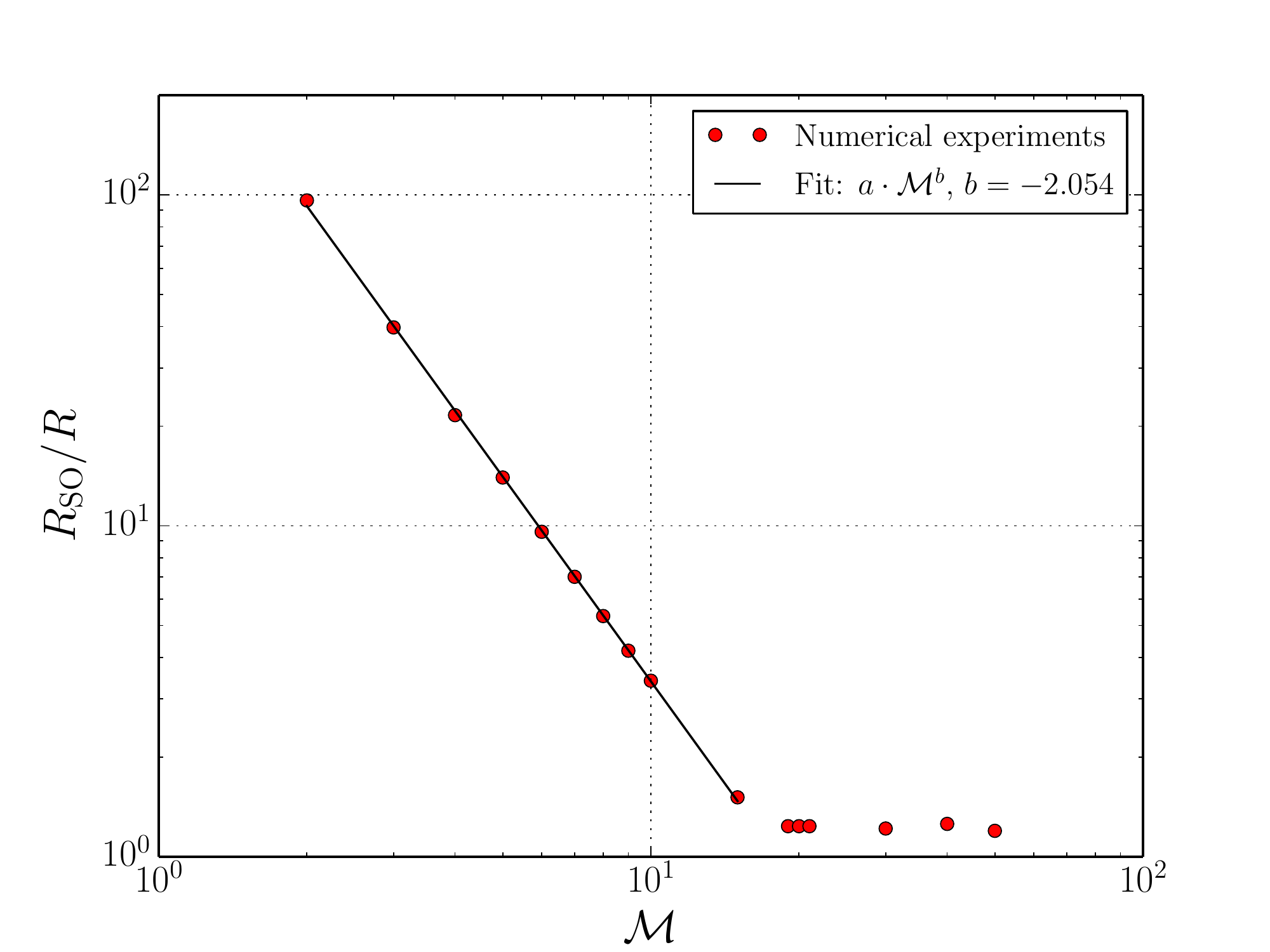} \\
\includegraphics[width=0.48\textwidth]{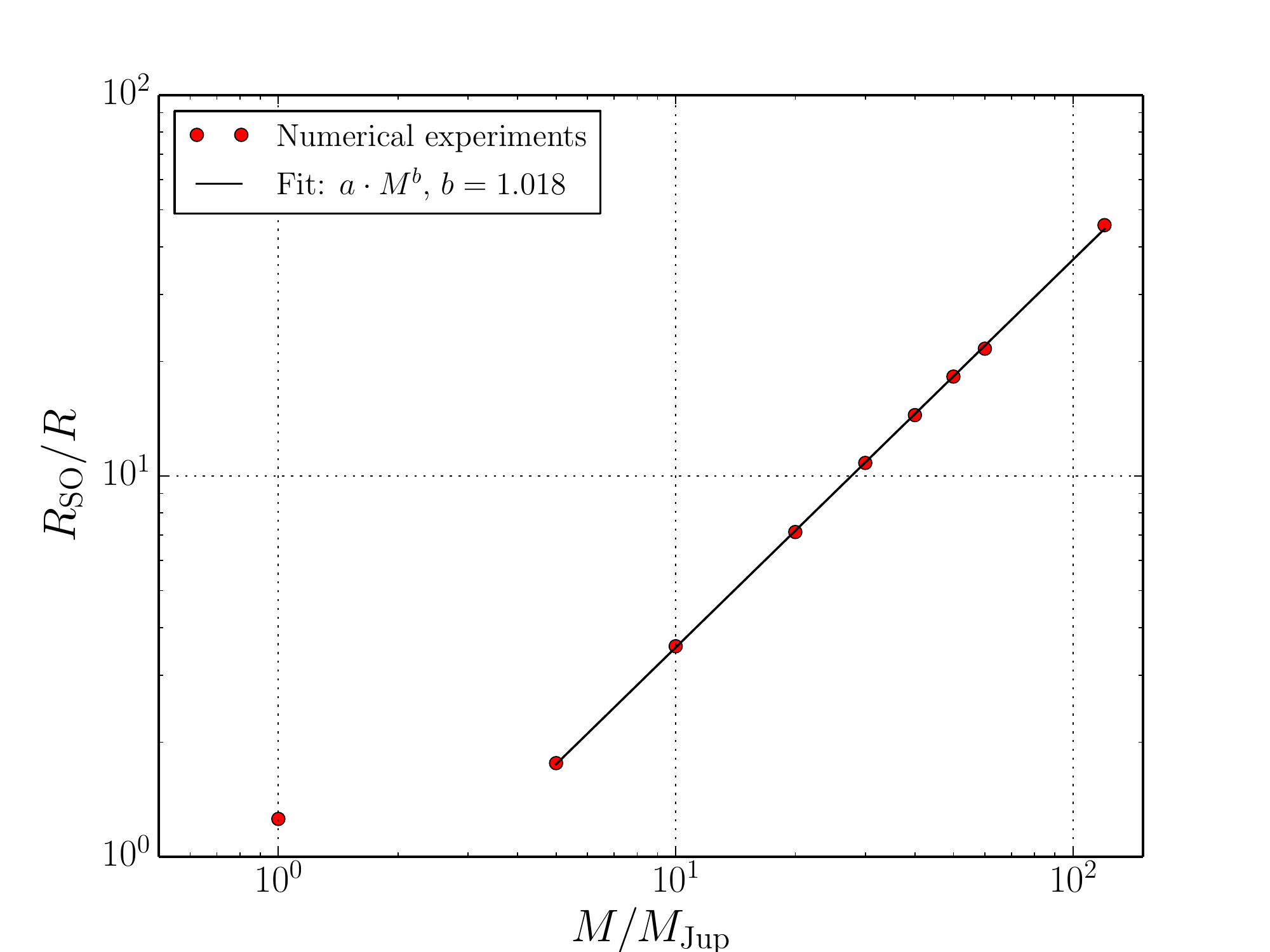}
\end{center}
\caption{
Correlation of the shock's stand-off distance $R_\mathrm{SO}$ with object
velocity or Mach number $\mathcal{M}$ respectively (top panel) and with object
mass $M$ (bottom panel).  Solid lines denote linear fits in the log-log plane.
}
\label{fig:rso-dependencies}
\end{figure}

Fig.~\ref{fig:radial-drag} seems to indicate that $R_\mathrm{SO}$ is directly
proportional to the accretion radius $R_\mathrm{A}$.  Since $R_\mathrm{A}$
scales as $\propto M/V_\infty^2$ we check these scaling laws for the stand-off
distance as well.  We use a series of simulations where we vary systematically
the mass $M$ of the object and its velocity $V_\infty$.  Simulation parameters
are given in Table~\ref{tab:sims}, series ``M2'' and ``V4''.  The results are
displayed in Fig.~\ref{fig:rso-dependencies}.
Both panels show least-square fit lines that support the fact that
$R_\mathrm{SO} \propto M/\mathcal{M}^2$.  For larger $\mathcal{M}$ the stand-off
distance becomes smaller according to $R_\mathrm{SO} \propto {\mathcal{M}^{-2}}$
as displayed in the top panel of Fig.~\ref{fig:rso-dependencies}.  For fixed
$\mathcal{M}$, lowering the object mass yields a smaller stand-off distance in
agreement with $R_\mathrm{SO} \propto M$ as shown in the bottom panel of
Fig.~\ref{fig:rso-dependencies}.  This outcome confirms the scaling laws as long
as the geometrical extent of the body $R$ is small compared to $R_\mathrm{SO}$.
For high Mach shocks and for low body masses, respectively, the stand-off
distance cannot decrease to smaller radii  and approaches the size of the body
$R$ instead.  In all other cases, $R_\mathrm{SO}$ is directly proportional to
$R_\mathrm{A}$:
\begin{equation}
R_\mathrm{SO} \propto R_\mathrm{A} \, .
\end{equation}
An important difference between both radii is given by the fact that the
stand-off distance depends on the thermodynamics of the system.  As a
consequence, we study the impact of the adiabatic index $\gamma$ on the
stand-off distance and the final dynamical friction term in
Sect.~\ref{sect:gamma} below.

\begin{figure}[htbp]
\begin{center}
\includegraphics[width=0.48\textwidth]{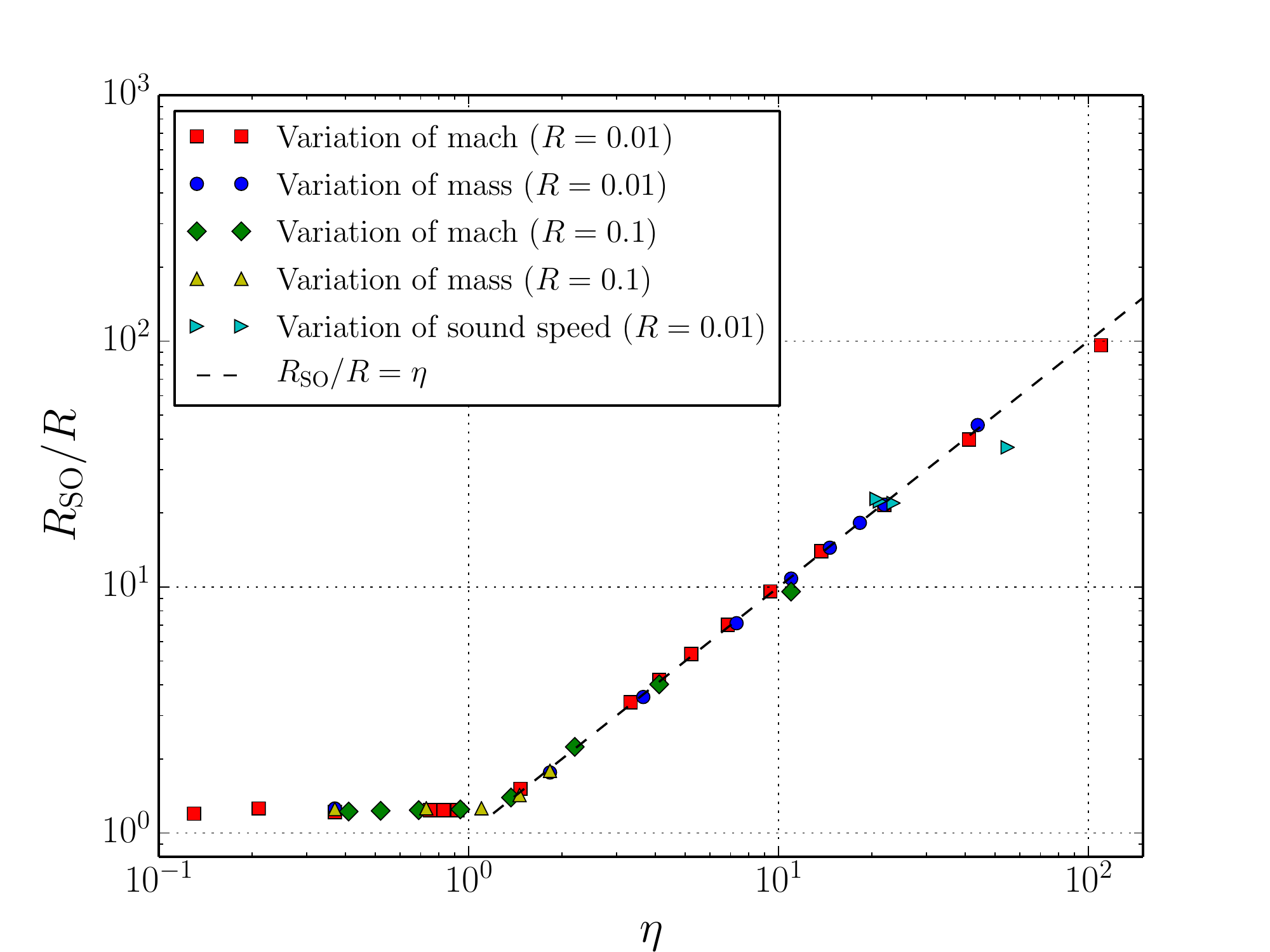} \\
\includegraphics[width=0.48\textwidth]{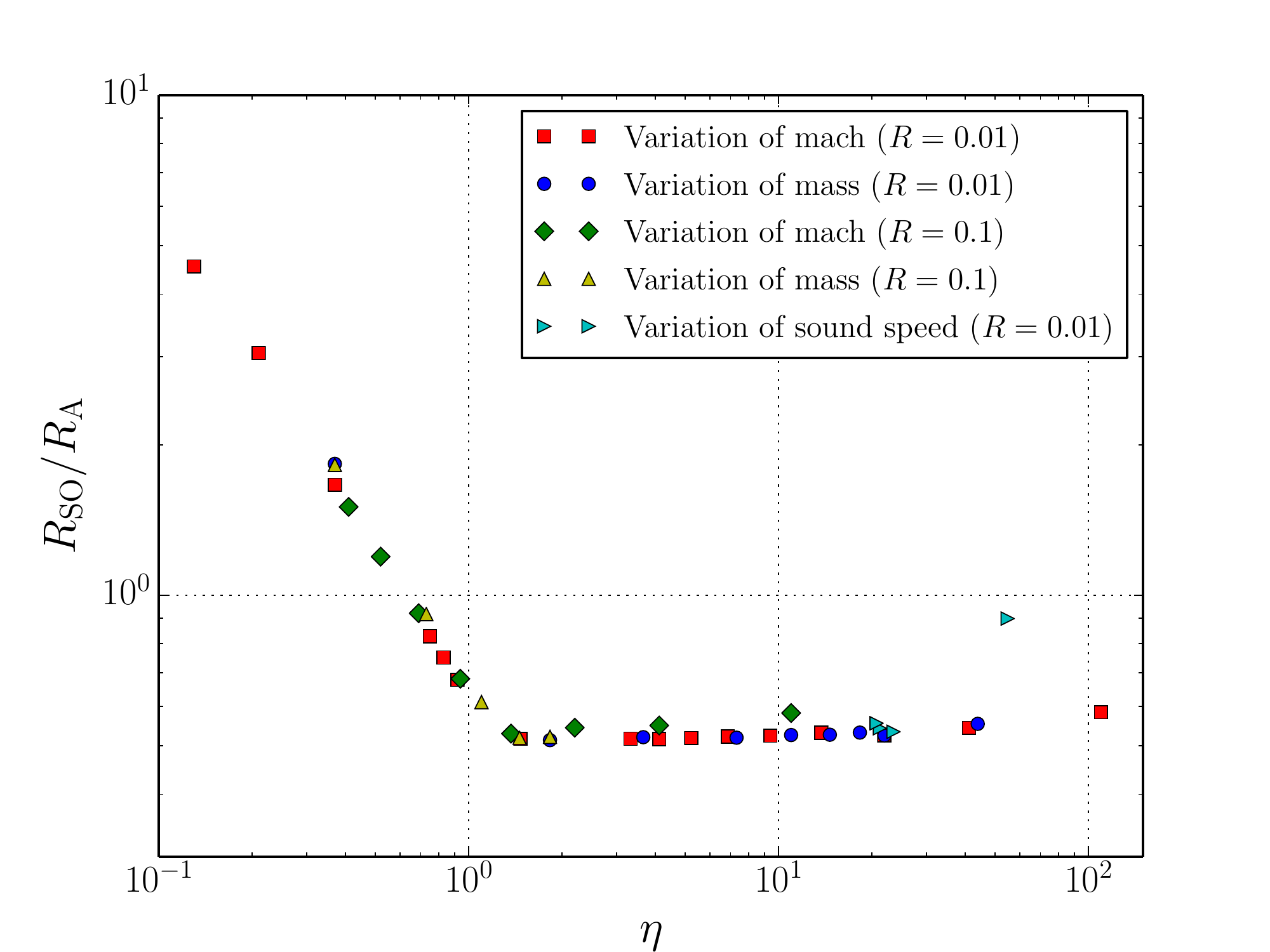}
\end{center}
\caption{
Correlation of the shock's stand-off distance $R_\mathrm{SO}$ with the
non-linearity parameter $\eta$, as given in Eq.~\eqref{eq:eta}, for various
model sequences.  In the top panel, $R_\mathrm{SO}$ is normalized by the object
radius $R$.  For $\eta >1$, a dashed line is superimposed, representing the
relation $R_\mathrm{SO}/R = \eta$.  In the bottom panel, $R_\mathrm{SO}$ is
normalized by the corresponding accretion radius.
} 
\label{fig:eta-dependence-g1.7}
\end{figure}

To investigate the proportionality between the two radii further, we perform
additional simulations varying the mass, radius, and velocity of the object as
well as the soundspeed of the medium.  Detailed simulation parameters are given
in Table~\ref{tab:sims}.  To find a useful general expression for the stand-off
distance $R_\mathrm{SO}$ we utilize the non-linearity parameter
\begin{equation}\label{eq:eta}
\eta = \frac{G M}{(\mathcal{M}^2 - 1)c_\infty^2 R}
= \frac{1}{2} ~ \frac{\mathcal{M}^2}{\mathcal{M}^2 - 1} \, \frac{R_\mathrm{A}}{R} \, ,
\end{equation}
as introduced by \citet{2009ApJ...703.1278K}.  In
Fig.~\ref{fig:eta-dependence-g1.7} we show the stand-off distance
$R_\mathrm{SO}$ as well as the ratio of $R_\mathrm{SO} / R_\mathrm{A}$ as
function of the non-linearity parameter $\eta$. 
As seen in the top panel, for $\eta > 1$ (larger $M$, lower $\mathcal{M}$) we
find the scaling $R_\mathrm{SO}/R \propto \eta$, with an unknown proportional
constant close to unity as indicated by the added dashed line.  This is in good
agreement with the results of \citet{2009ApJ...703.1278K}.  The largest
deviation from this trend is seen in the rightmost green triangle, which belongs
to a simulation in the weakly supersonic regime $\mathcal{M} =1.25$.  A spatial
analysis of the final configuration of this model shows that behind the shock
front and around the object the gaseous medium is not in hydrostatic equilibrium
-- as for the supersonic cases -- but displays small residual motions that did
not disappear even in long-term runs.  So, we expect deviations from the
following relation \eqref{eq:Rso-Mach} for non-supersonic flows, a regime that
is not part of the present study and will have to be explored in future studies.
For comparison, we denote that the rightmost red square belongs to a simulation
with $\mathcal{M}=2$, but even larger $\eta$ than the green triangle run, and
matches the found relationship much better.

For $\eta < 1$ (larger $\mathcal{M}$, lower $M$) we enter the regime where the
stand-off distance is limited by the size of the object $R$, and therefore the
ratio $R_\mathrm{SO}/R$ remains constant.  For this regime of low $\eta$,
\citet{2009ApJ...703.1278K} find a different scaling due to the fact that they
used a smoothed gravitational potential to model their perturber instead of the
solid surface used in our study.  Using the definition of $\eta$
(Eq.~\eqref{eq:eta}), the proportionality $R_\mathrm{SO}/R \propto \eta$ implies
\begin{equation} \label{eq:Rso-Mach}
R_\mathrm{SO} \propto \frac{1}{2} \, \frac{\mathcal{M}^2}{\mathcal{M}^2 - 1} ~ R_\mathrm{A} \mbox{~~for  } R_\mathrm{SO} \gg R\, .
\end{equation}
Relation \eqref{eq:Rso-Mach} represents a suitable expression for the stand-off
distance.  As pointed out above, the proportionality constant for the
simulations performed so far seems to be very close to unity.  But the relation
still lacks the effect of different equations of state, i.e.~any dependence on
the adiabatic index $\gamma$.  We investigate this dependence in detail in the
following section.

The bottom panel of Fig.~\ref{fig:eta-dependence-g1.7} just confirms for
completeness again the proportionality between the stand-off distance and the
accretion radius for different Mach numbers, object masses, and soundspeeds in
the regime $R_\mathrm{SO} \gg R$.  We checked in addition that a change in the
size $R$ of the moving body does not influence the stand-off distance as long as
$R$ remains substantially smaller than $R_\mathrm{SO}$. Hence, the
curves for $R=0.1$ and $R=0.01$ are identical.

\subsection{The maximum extent of the wake} \label{sect:smax}
The maximum extent of the wake past the body determines the size of the
perturbed region, which will contribute to the total dynamical friction via its
gravitational pull.  In Eq.~\eqref{eq:fdrag0}, the maximum extent of the wake
enters as the outer radius of the integral $s_\mathrm{max}$ in the
Coulomb-Logarithm $C_A$:
\begin{equation} \label{eq:smax}
F_\mathrm{DF} \propto \ln({s_\mathrm{max}} / R_\mathrm{SO}).
\end{equation} 

For an infinite medium, the size of the wake $s_\mathrm{max}$ is given by the
duration of the movement of the perturbing object.  For a body starting its
journey at $t=0$ with a constant velocity $V_\infty$, the time-dependent size of
the wake $s_\mathrm{max}$ at a later time $t$ is given by the distance travelled
$s_\mathrm{max}(t) = V_\infty t$.

The relation \eqref{eq:smax} implies that the dynamical friction increases as
long as the body is moving through the medium, leading to a larger and larger
wake.  In principal, the dynamical friction seems to approach an infinitely high
force for an infinitely long traveling object.  In practice, the dynamical
friction can not grow infinitely, even not for an infinitely large medium,
because the dynamical friction acting on the body leads to a slow-down of the
object on a dynamical timescale of the gas-body interaction
\citep{2001MNRAS.322...67S}.  The dynamical timescale of dynamical
friction is given as the ratio of the initial momentum of the moving object and
the acting force $t_\mathrm{DF} = M V_\infty / F_\mathrm{DF}$.  Hence, the
velocity of the perturber monotonically decreases in time until the body is at
rest.  As a consequence, the dynamical friction first increases (due to a
growing extent of the wake) up to a maximum value and decreases afterwards (due
to the slow-down of the body).

Here, we model the evolution of the system on timescales much smaller than the
timescale for the slow-down of the body to extract the acting dynamical friction
as a function of the relevant initial system parameters.  As a next step, we
check the scaling of the dynamical friction force with the extent of the wake.
We compute the resulting dynamical friction by numerically integrating the
gravitational pull of the forming wake onto the body up to different radii
within the computational domain.  As a result, we obtain the individual
contributions of the wake at each radius.  For further analysis, we extended the
size of the computational domain to $R_\mathrm{domain} = 25 \mbox{R}_\odot$
(model C), other simulation parameter correspond to the fiducial case (model F)
given in Table~\ref{tab:sims}.

\begin{figure}
    \begin{center}
    \includegraphics[width=0.49\textwidth]{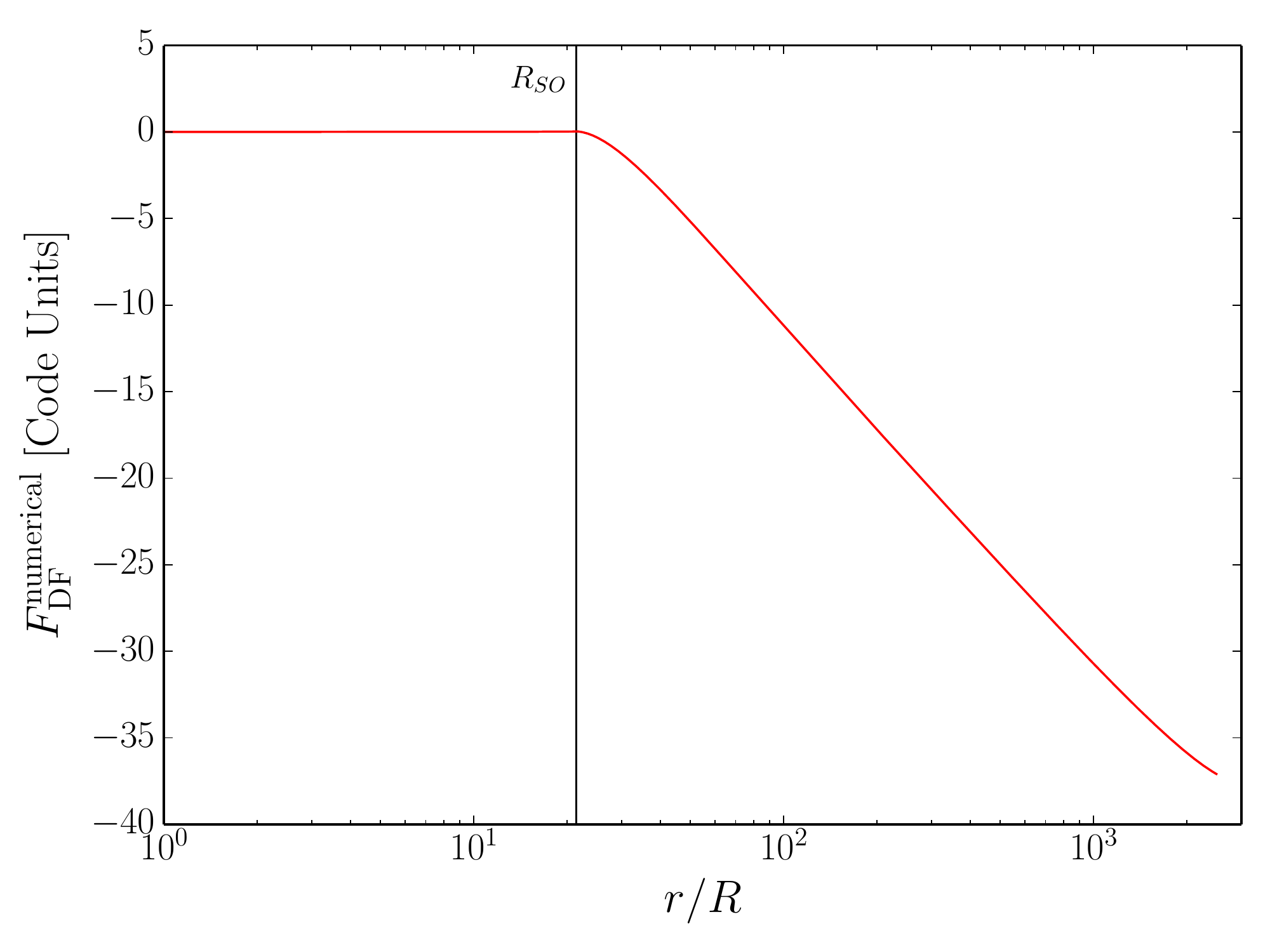}
    \end{center}
    \caption{Scaling of dynamical friction with the size of the forming wake
        for model C in Table~\ref{tab:sims}.}
    \label{fig:smax}
\end{figure}

As shown in Fig.\ref{fig:smax}, the scaling of the numerically determined drag
force with the extent of the wake follows the expected logarithmic relation.
In the following, simulations make use of an arbitrary size of the computational
domain, which just should be much larger than the stand-off distance of the
shock.  The finally acting dynamical friction can then be determined using the
scaling relation Eq.~\eqref{eq:smax}.  We use an outer radius of the
computational domain of either $100~R$ or $1000~R$, depending on the specified
radius $R$ of the moving body.

\subsection{The mass of the perturber} \label{sect:scaleM}
The dimensional part of the dynamical friction formula -- as written in
Eqs.~\eqref{eq:fdrag0} and \eqref{eq:drag} -- is expected to scale with the mass
of the object squared:
\begin{equation} \label{eq:scaleM}
    F_\mathrm{DF} / C_A \propto M^2 \,.
\end{equation} 
The physical explanation for this scaling relation is that the mass of the
object first causes the accumulation of high density in the wake (a
gravitational back-reaction, which scales linearly with the mass of the object),
and secondly, the gravitational pull of the wake onto the object denotes a
gravitational interaction, which again scales linearly with the mass of the
object.  But this argument is based on a linearization of the equations to
relate the forming wake to a linear scaling with $M$.

\begin{figure}[htbp]
    \begin{center}
    \includegraphics[width=0.49\textwidth]{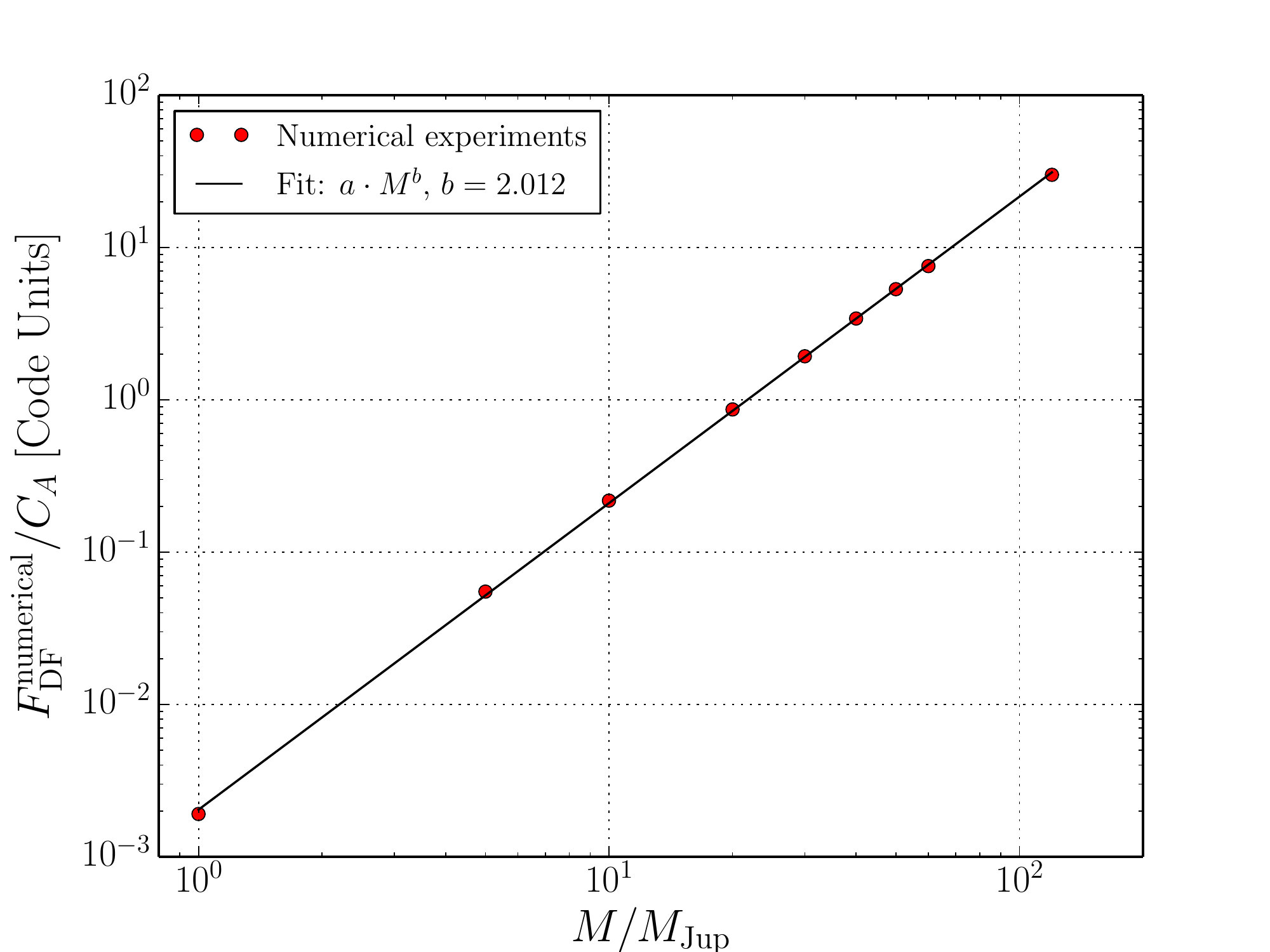}
    \end{center}
    \caption{Correlation of the ratio of dynamical friction and Coulomb
        logarithm with the mass of the perturber.} 
    \label{fig:scaleM}
\end{figure}

Within our numerical framework, we can compute the total dependence of the
dynamical friction on the mass, and can further check, if the dependence on $M$
can properly be split into a scaling with $M^2$ in the dimensional part and its
further dependence within the dimensionless Coulomb logarithm $C_A$.  We have
performed numerical simulations for various masses $M$ of the perturber.
Simulation parameters are given in Table~\ref{tab:sims}, series ``M2''.  The
resulting scaling of the ratio of dynamical friction and Coulomb logarithm is 
shown in Fig.~\ref{fig:scaleM}.
\newline
The numerical experiments confirm the expected scaling very accurately.

\subsection{The velocity of the perturber} \label{sect:scaleV}
The dimensional part of the dynamical friction formula -- as written in
Eqs.~\eqref{eq:fdrag0} and \eqref{eq:drag} -- is expected to scale with the
inverse of the velocity of the object squared:
\begin{equation} \label{eq:scaleV}
    F_\mathrm{DF} / C_A \propto V_\infty^{-2}
\end{equation}
This scaling can only be true for supersonic motion and actually denotes a
peculiarity of the dynamical friction, because usually hydrodynamical drag
forces, as denoted in Eq.~\eqref{eq:drag}, scale with the square of the
velocity, $V_\infty^2$.  But in the case of dynamical friction, a faster
body is already further away from the high-density wake, once it has formed.
The distance of the body to the forming wake scales linearly with its velocity
and the gravitational pull scales with the inverse of the distance squared.

As shown above, this inverse scaling with $V_\infty^2$ of the dynamical
friction force, can be understood in terms of an effective cross section that
is given by the accretion radius, $R_\mathrm{acc}$, that decreases with higher
velocities of the body.  This gives rise to the scaling of the dynamical
friction with the inverse of the velocity of the object squared.

\begin{figure}[htbp]
    \begin{center}
    \includegraphics[width=0.49\textwidth]{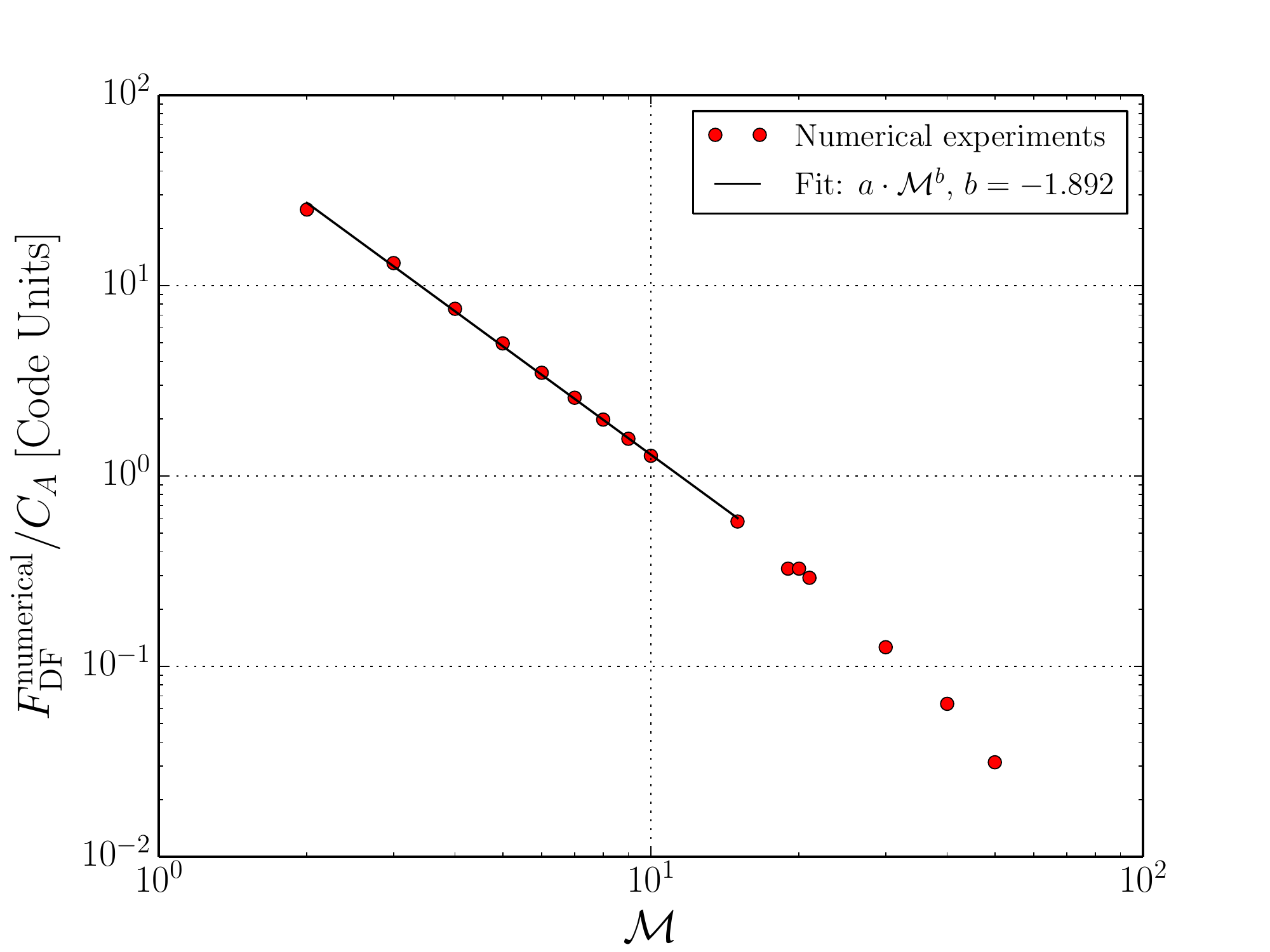}
    \end{center}
    \caption{ Correlation of the ratio of dynamical friction and 
      Coulomb logarithm with the velocity of the perturber.} 
    \label{fig:scaleV}
\end{figure}

We have performed numerical simulations for various velocities $V_\infty$ of the
perturber.  Simulation parameters are given in Table~\ref{tab:sims}, series
``V4''.  The resulting scaling of the ratio of dynamical friction and drag
coefficient is shown in Fig.~\ref{fig:scaleV}.
Again, the numerical experiments confirm the expected scaling.  Deviations from
the scaling law occur for very high Mach numbers, because the stand-off distance
cannot shrink to arbitrarily small values and approaches the geometrical radius
of the body instead, cp.~discussion in Sect.~\ref{sect:stand-off} for details,
especially Fig.~\ref{fig:rso-dependencies}.

\subsection{The adiabatic index}
\label{sect:gamma}
All the simulations presented in the previous sections use an adiabatic index of
$\gamma = 5/3$.  This was also the one and only value for the adiabatic index
investigated in \citet{2009ApJ...703.1278K}.  But in fact, the compression in
the shocked region around the object as well as the compression in the wake
behind the moving body depends on the equation of state in use, and hence, on the
value of the adiabatic index.  In an extreme case, for very large values of the
adiabatic index, the gaseous medium becomes incompressible for example, because any
compression (increase in density) causes the pressure to increase infinitely
strong, which causes an infinitely strong pressure force, and the medium
directly relaxes towards an iso-density morphology again.

To further investigate the generality of our results so far we perform a variety
of additional model sequences using different adiabatic indices.  Detailed setup
parameters of these simulation series are given in Table~\ref{tab:sims}.

\begin{figure}[htbp]
    \begin{center}
    \includegraphics[width=0.49\textwidth]{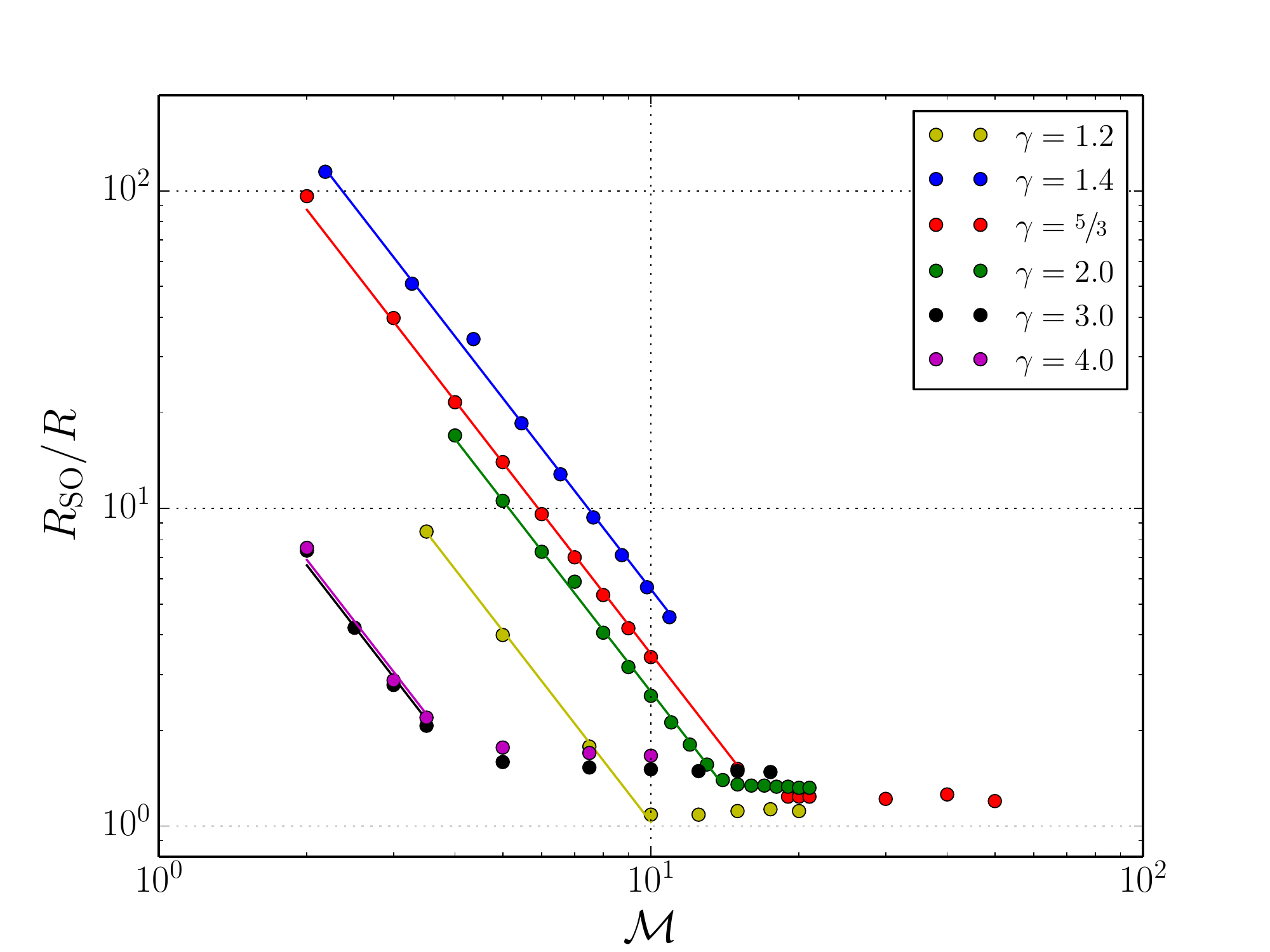} 
    \end{center}
    \caption{Correlation of the shock's stand-off distance with the Mach number
        for different values of the adiabatic index $\gamma$.  Dots denote
        results from the numerical simulations.  The colored lines represent the
        scaling of $R_\mathrm{SO}/R \propto \mathcal{M}^{-2}$.}
    \label{fig:rso-mach-gamma}
\end{figure}

The correlation of the stand-off distance with the velocity of the perturber is
shown in Fig.~\ref{fig:rso-mach-gamma}, but now for a variety of different
values of the adiabatic index $\gamma$. 
For large Mach numbers $\mathcal{M}$, the stand-off distance approaches the
geometrical radius of the object.  For lower $\mathcal{M}$, as indicated by the
colored lines, all simulation series confirm the scaling law of $R_\mathrm{SO}/R
\propto \mathcal{M}^{-2}$ as previously determined in
Fig.~\ref{fig:rso-dependencies}.

\begin{figure}[htbp]
    \begin{center}
    \includegraphics[width=0.49\textwidth]{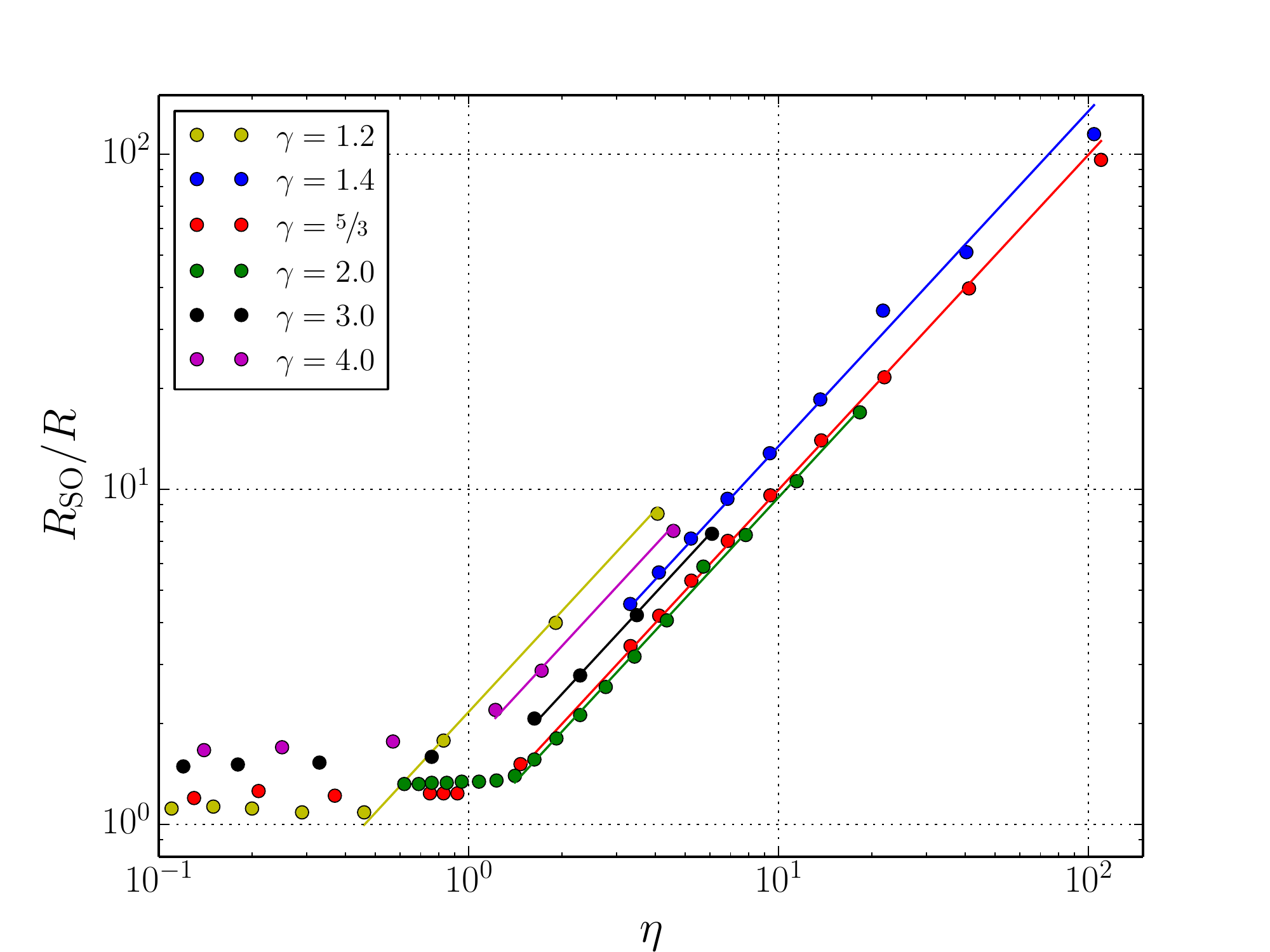} \\
    \includegraphics[width=0.49\textwidth]{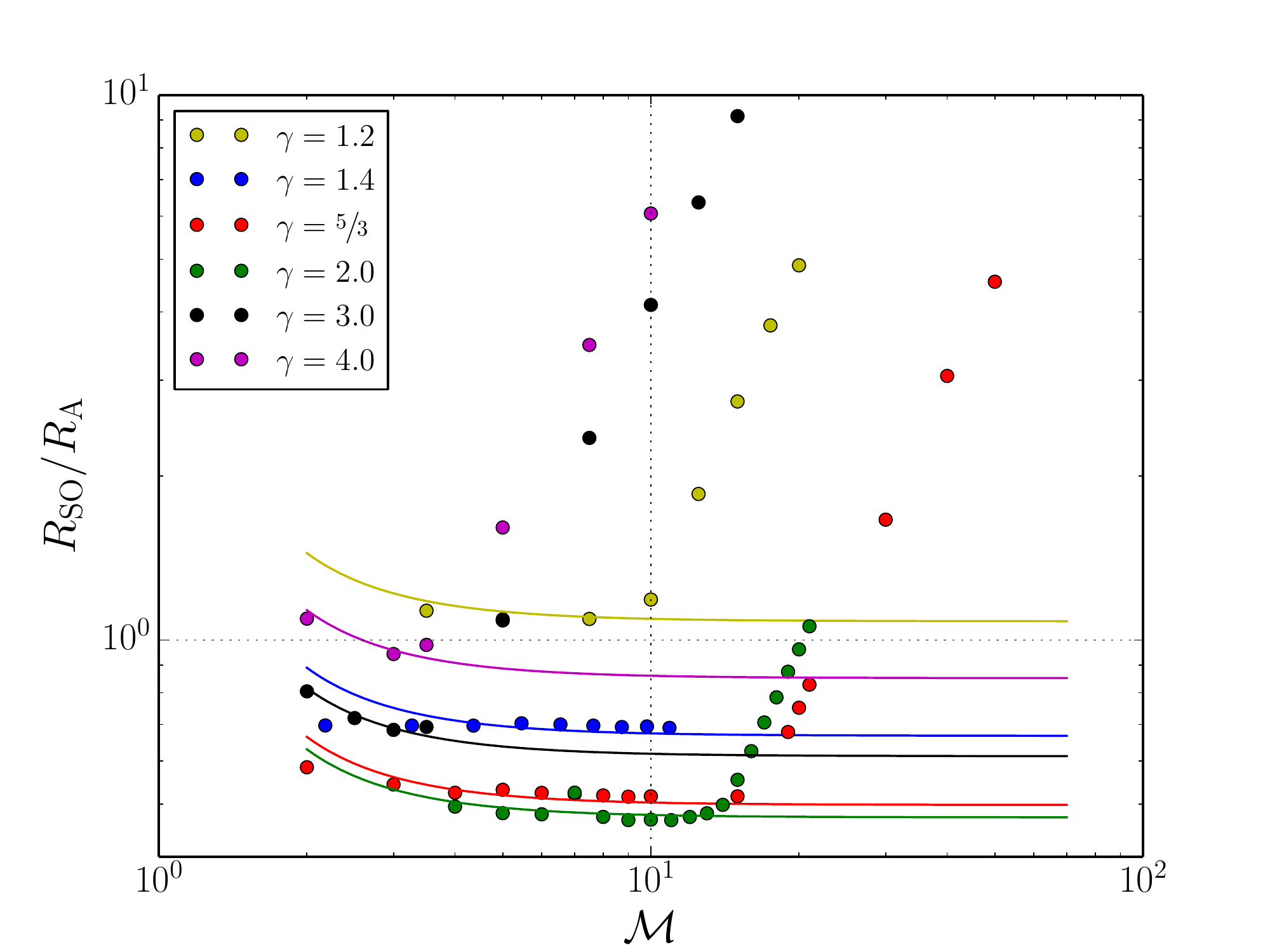}
    \end{center}
    \caption{Correlation of the shock's stand-off distance with the
        non-linearity parameter $\eta$ given in Eq.~\eqref{eq:eta} (top panel)
        and the Mach number (bottom panel). The top panel is analogous to the
        top panel of Fig.~\ref{fig:eta-dependence-g1.7}, but now with results
        for different values of the adiabatic index $\gamma$.  Dots denote
        results from the numerical simulations.  Colored lines represent the
        relations discussed in the main text.}  
    \label{fig:eta-dependence-g-multi}
\end{figure}

In Fig.~\ref{fig:eta-dependence-g-multi} top panel, the correlation of the
stand-off distance with the non-linearity parameter $\eta$ is shown for
different values of $\gamma$.
Colored lines indicate linear fits in the log-log plane of the plot.  Since we
found $R_\mathrm{SO}/R \propto \eta$ for $\gamma = \sfrac{5}{3}$ in agreement
with \citet{2009ApJ...703.1278K}, we assume the following, more general
relation:
\begin{equation} \label{eq:Rso-R-general}
    \frac{R_\mathrm{SO}}{R} = g(\gamma) \, \eta \, ,
\end{equation}
where the factor $g(\gamma)$ describes the dependence on the adiabatic index
$\gamma$.  Specific values of this function are obtained according to the linear
fits to the numerical experiments, namely

\begin{tabular}{l | c c c c c c}
$\gamma$ 	& 1.2 & 1.4 & 5/3 & 2.0 & 3.0 & 4.0 \\
\hline
$g(\gamma$)	& 2.17 & 1.33 & 1.00  & 0.95 & 1.22  & 1.70 
\end{tabular}
\newline

\noindent
For large values of the adiabatic index, namely $\gamma = 3$ and $4$, the
accuracy of the fit values has to be taken with care, because the fits rely on
three to four data points only.

In the bottom panel of Fig.~\ref{fig:eta-dependence-g-multi}, the stand-off
distance is shown in units of the accretion radius as a function of the Mach
number.  Also here, the previously determined correlation between the stand-off
distance and the accretion radius (Eq.~\eqref{eq:Rso-Mach}) can be generalized
to include the dependence on the adiabatic index $\gamma$ via:
\begin{equation} \label{eq:Rso-Mach-gen}
    R_\mathrm{SO} = \frac{g(\gamma)}{2} 
                    \, \frac{\mathcal{M}^2}{\mathcal{M}^2 - 1} 
                    \, R_\mathrm{A} \, .
\end{equation}
This function is presented as colored lines in the bottom panel of
Fig.~\ref{fig:eta-dependence-g-multi}.  Values for $g(\gamma)$ were used as
given above.

\begin{figure}[htbp]
    \begin{center}
    \includegraphics[width=0.49\textwidth]{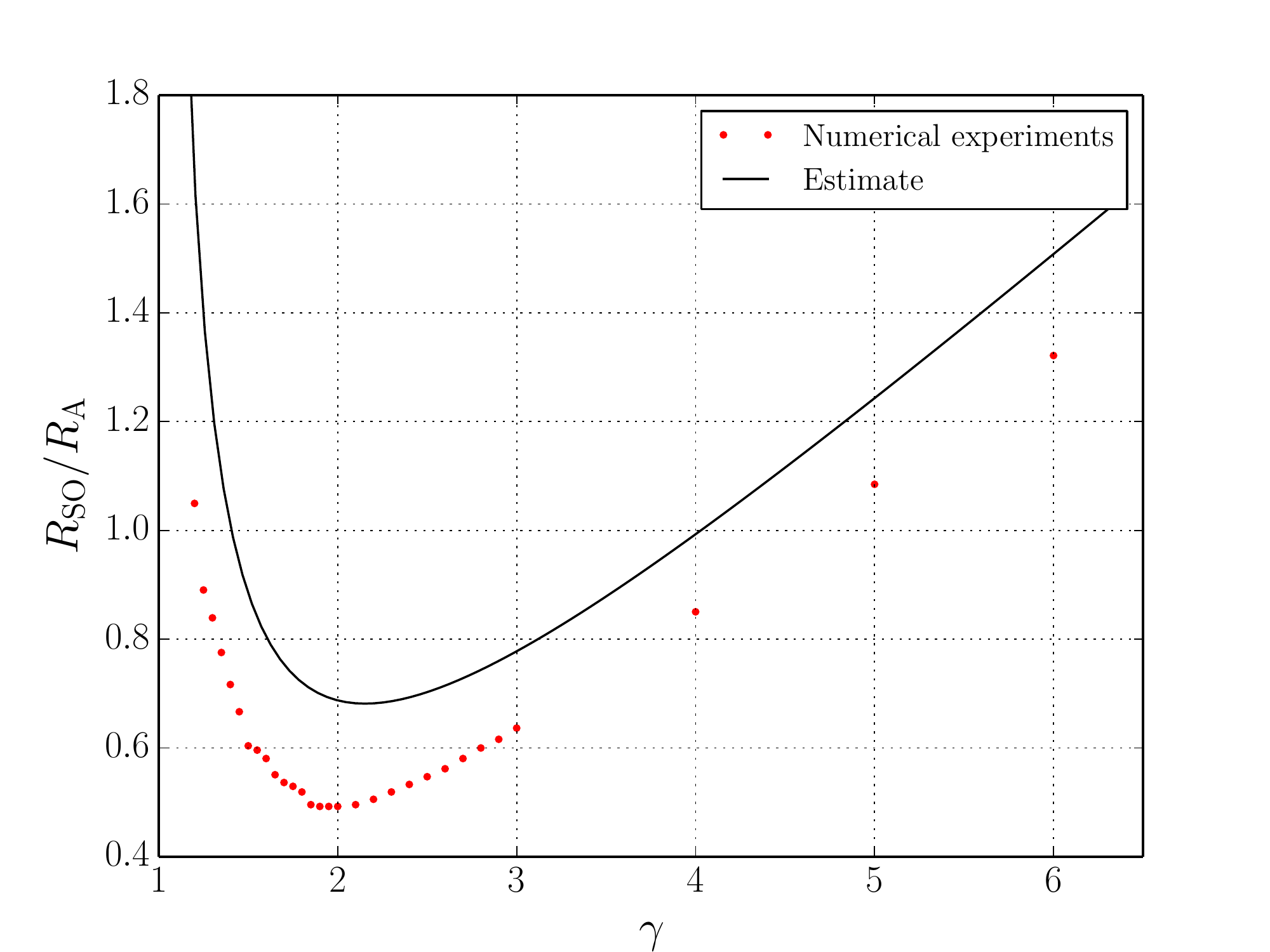}
    \end{center}
    \caption{ Stand-off distance in units of accretion radius
        $R_\mathrm{SO}/R_\mathrm{A}$ as a function of the adiabatic index
        $\gamma$.  The black line denotes the analytically estimated upper limit
        given in Eq.~\eqref{eq:xso_base}.}
    \label{fig:gam-rso}
\end{figure}

In appendix~\ref{app:stand-off}, we derive an analytical estimate of the
$\gamma$-dependence of the stand-off distance:
\begin{equation} \label{eq:xso_base}
    \frac{R_\mathrm{SO}}{R_\mathrm{A}} \lesssim \frac{((\gamma) + 1)^2}{4(\gamma)} \left[1 - \left(\frac{((\gamma)-1)}{(\gamma)+1}\right)^{(\gamma)-1}\right]^{-1} - 1
\end{equation}
This relation is shown in comparison to a simulation series of varying adiabatic
index in Fig.~\ref{fig:gam-rso}.
The estimate does not allow for an exact determination, but it denotes an upper
limit of the stand-off distance.  Most importantly, the estimate follows the
trend of the $\gamma$-dependence when compared to the numerical experiments.

Due to the similarity to the $\gamma$-dependence, we choose the analytical
estimate as the basis function of a fit to the numerical data, but allow now for
a shift in $\gamma \rightarrow \gamma + a$ and in $R_\mathrm{SO}/R_\mathrm{A}
\rightarrow R_\mathrm{SO}/R_\mathrm{A} + b$:
\begin{equation}
\label{eq:xso_fit}
\frac{R_\mathrm{SO}}{R_\mathrm{A}} \approx \frac{((\gamma+a) + 1)^2}{4(\gamma+a)} 
    \left[1 - \left(\frac{((\gamma+a)-1)}{(\gamma+a)+1}\right)^{(\gamma+a)
    -1}\right]^{-1} - 1 - b
\end{equation}
with $a$ and $b$ as free fitting parameters.  A least square fit of this
function to the numerical data yields $a = 0.1$ and $b = 0.18$.

Using Eq.~\eqref{eq:Rso-Mach-gen} for the $\mathcal{M} = 4.0$ numerical data in
combination with the fitted function Eq.~\eqref{eq:xso_fit} allows to give an
approximate function of $g(\gamma)$:
\begin{equation} \label{eq:g_factor}
g(\gamma) \approx \frac{15}{8} \left\{ \frac{((\gamma+a) + 1)^2}{4(\gamma+a)} 
    \left[1 - \left(\frac{((\gamma+a)-1)}{(\gamma+a)+1}\right)^{(\gamma+a)
	-1}\right]^{-1}\negthickspace -1 - b \right\}
\end{equation}
with $a = 0.1$ and $b = 0.18$.

\begin{figure}
    \begin{center}
    \includegraphics[width=0.49\textwidth]{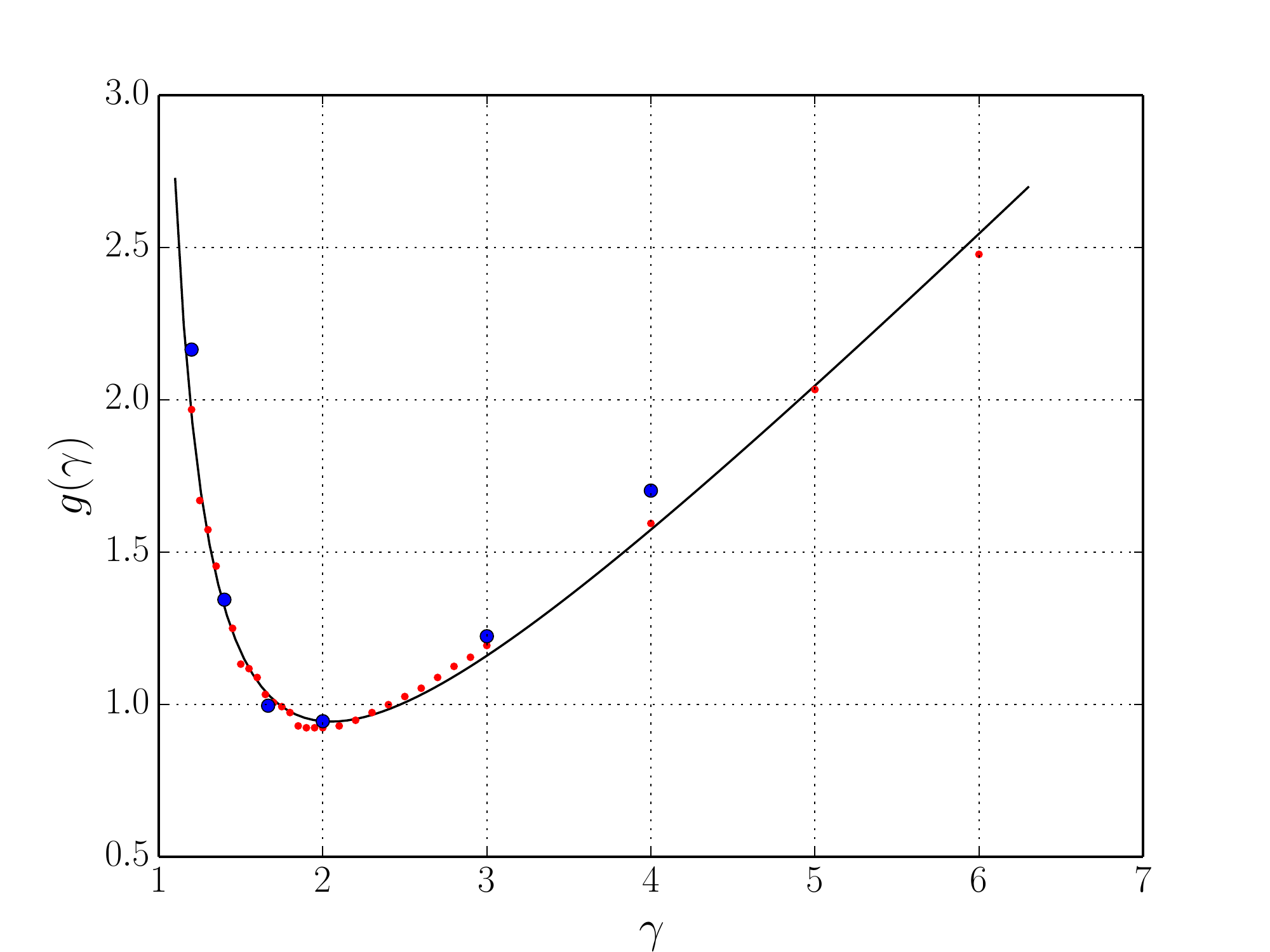} 
    \end{center}
    \caption{Comparison of the semi-analytical approximation of the function
        $g(\gamma)$, see Eq~\eqref{eq:g_factor}, shown as a black line with
        numerically obtained values.  Small red dots denote the simulation
        series ``G'' with varying adiabatic index; each red dot represents a
        single simulation.  Larger blue dots denote values derived by fitting
        the scaling of the stand-off distance with the non-linearity parameter
        for a variety of different simulation series shown in the top panel of
        Fig.~\ref{fig:eta-dependence-g-multi}; each blue dot represents such a
        fit value to a full simulation series.  See main text for details of the
        derivation.}
    \label{fig:g_factor}
\end{figure}

As a further check of the approximate function, we can compare this relation
with the tabular values above derived via Eq.~\eqref{eq:Rso-R-general}, see
Fig.~\ref{fig:g_factor}.
The approximate function $g(\gamma)$ for the overall $\gamma$-dependence gives
reasonable results in comparison to the numerical outcome.  As stated
previously, for large values of the adiabatic index, namely $\gamma = 3$ and
$4$, the accuracy of the fit values (big blue dots) has to be taken with care,
because the fits rely on three to four data points only.

\section{A new formula of dynamical friction}
\label{sect:dragforce}
As a final step, we combine  our obtained scaling law for the stand-off
distance, $R_\mathrm{SO}$, and the dependence on the adiabatic index, $\gamma$,
to formulate a new general expression for the dynamical friction of a
gravitating body of mass $M$ and radius $R$ moving supersonically with velocity
$V_\infty$ through a homogeneous gaseous medium of mass density $\rho_\infty$
and soundspeed $c_\infty$:
\begin{equation} \label{eq:fdrag-final}
F_\mathrm{DF} = 4\pi\rho_\infty \left(\frac{G M}{V_\infty} \right)^2 ~ \ln \left( \frac{s_\mathrm{max}}{R_\mathrm{SO}} \right).
\end{equation}

The determination of the drag force from the basic problem parameters only
requires an a priori knowledge of the stand-off distance.  As shown above, the
shock's stand-off distance $R_\mathrm{SO}$ can be approximated by
\begin{equation}
R_\mathrm{SO} = \frac{g(\gamma)}{2} ~ \frac{\mathcal{M}^2}{\mathcal{M}^2 - 1} ~ R_\mathrm{A},
\end{equation}
with $g(\gamma)$ given by Eq.~\eqref{eq:g_factor}.

Strictly speaking, the formula above for the stand-off distance is only valid
for a stand-off distance larger than the geometrical radius of the object.  In
the regime of large, low-mass bodies moving with high velocity, the stand-off
distance approaches the value of the geometrical radius instead, and one
can replace $s_\mathrm{min}$ by the object size $R$.

\begin{figure}[htbp]
    \begin{center}
    \includegraphics[width=0.49\textwidth]{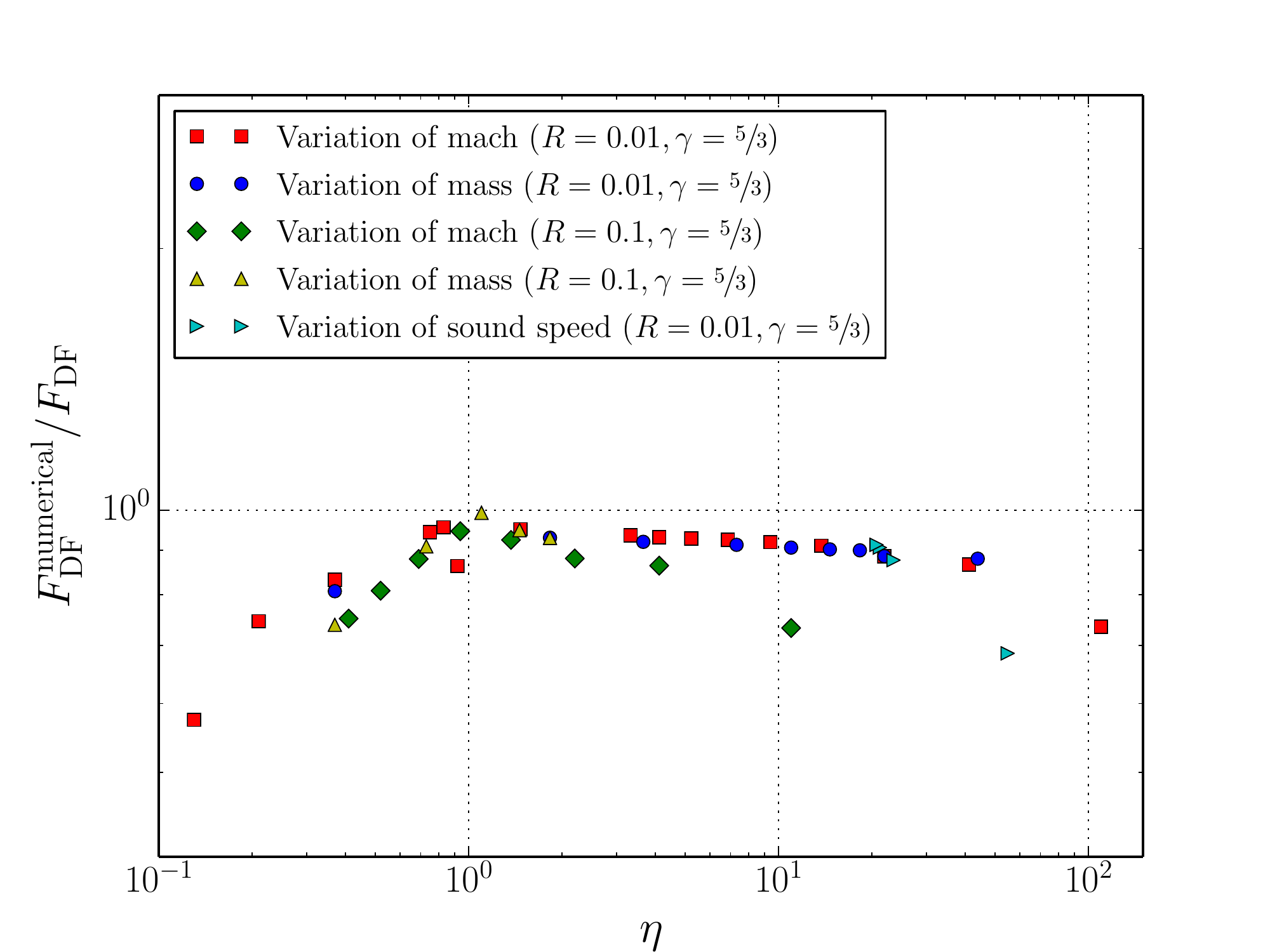} \\
    \includegraphics[width=0.49\textwidth]{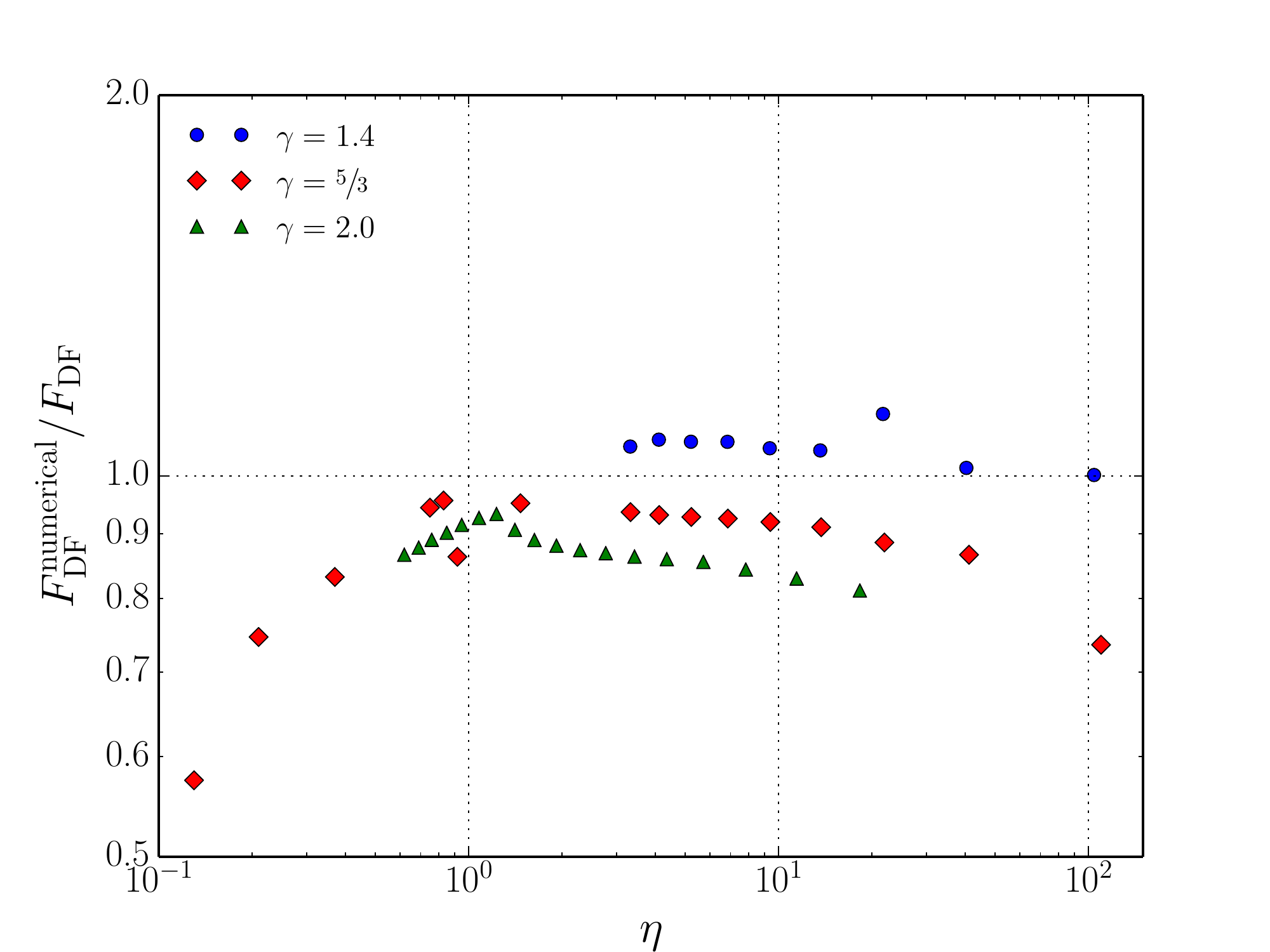}
    \end{center}
    \caption{Comparison of the newly derived approximate formula for dynamical
        friction $F_\mathrm{DF}$ with data from various numerical experiments
        $F^\mathrm{numerical}_\mathrm{DF}$.  The corresponding ratio is given as function of
        the non-linearity parameter $\eta$.  Top panel shows results from a
        variety of simulations with adiabatic index $\gamma = 5/3$, varying the
        Mach number, soundspeed, and object mass.  Bottom panel shows results
        from a variety of simulations with varying Mach number for different
        adiabatic indices as labeled.} 
    \label{fig:r_so-r_a}
\end{figure}

We check the overall accuracy of this approximate determination of dynamical
friction in direct comparison to the various numerical experiments performed
herein, covering a broad parameter space in terms of the dimensionless
non-linearity parameter $\eta = [2 ~ (\mathcal{M}^2 - 1) / \mathcal{M}^2 ~ R /
R_\mathrm{A}]^{-1}$.  The result of this comparison is shown as the ratio of the
force in the experiments and the approximate formula in Fig.~\ref{fig:r_so-r_a}. 
The approximate formula can be used as a convenient tool to estimate the drag
force from the basic parameters in reasonable accuracy.  Especially in the
regime of $1 \le \eta \le 50$ the estimated values match the numerical
experiments.  Largest deviations are observed in the marginally supersonic
regime ($\mathcal{M}$ close to unity, denoted by the rightmost greenish
triangle) and in cases, where the shock's stand-off distance approaches the size
of the object.  The latter effect is visible when comparing the numerical
outcome in the regime of large $\eta$ for simulations with a larger (green
diamonds) and smaller object radii (red squares) in the top panel of
Fig.~\ref{fig:r_so-r_a}.

\begin{figure}[htbp]
    \begin{center}
    \includegraphics[width=0.49\textwidth]{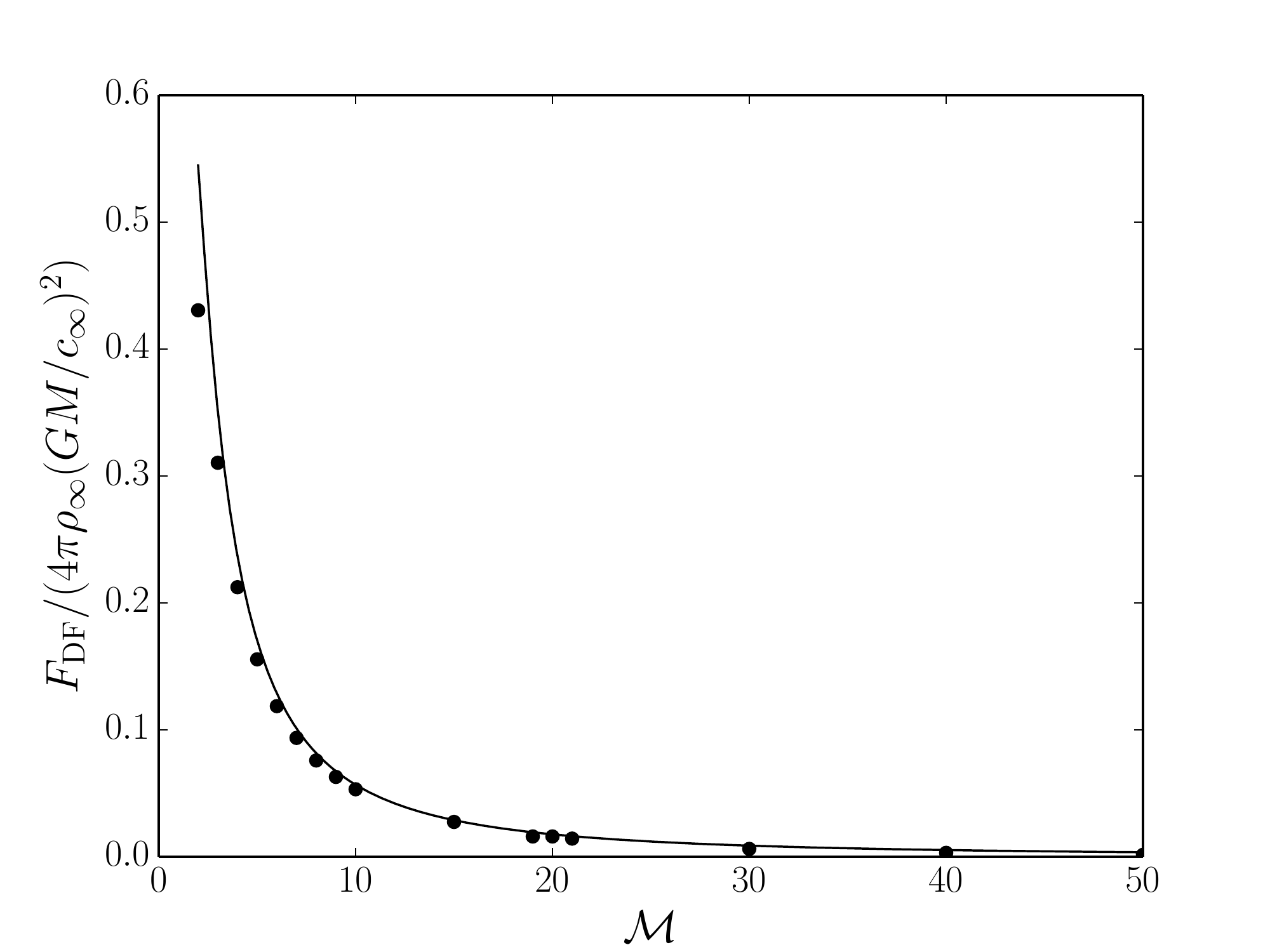}
    \end{center}
    \caption{The normalized drag force as a function of the Mach number.
             The solid line resembles our newly derived formula as given in Eq.~(\ref{eq:fdrag-final}).
             The numerical data are given by the black dots, for which we 
             we used simulation series V4.}
    \label{fig:df-factor-mach}
\end{figure}
To compare how well our newly derived formula approximates the drag force,
$F_\mathrm{DF}$, we plot in Fig.~\ref{fig:df-factor-mach} the numerically
calculated drag force divided by the factor $4\pi\rho_\infty \left( G M /
c_\infty\right)^2$ for varying Mach numbers of the perturber as black dots. The
relation
\begin{equation}
    F_\mathrm{DF} / (4\pi\rho_\infty \left( G M / c_\infty \right)^2) 
	= C_A / \mathcal{M}^2
\end{equation}
is shown as solid black line, where we use $C_A =
\ln(s_\mathrm{max}/R_\mathrm{SO})$ with $R_\mathrm{SO}$ according to
equation~\eqref{eq:Rso-Mach-gen}.  For the maximum radius in $C_A$ we used
$r_{max} = 1000 R$.  As seen our approximative function for $C_A$ reflects the
numerical data very well.

\section{Summary} \label{sect:summary}
We derived a new formula for dynamical friction of a body moving with supersonic
speed through a homogeneous gaseous medium.  This formula was obtained by
following an ansatz of direct numerical modeling of the dynamical problem.

In 11 simulation series, including slightly more than 100 individual simulations
in total, we scanned the parameter space of the problem and derived the scaling
relations of the dynamical friction with the mass, the velocity, and radius of
the moving body, the gas mass density, its soundspeed, and the value of the
adiabatic index, as well as the maximum extent of the forming wake.

\subsection{Scaling relations}
As expected, the dimensional part of the dynamical friction formula as given in
Eq.~\eqref{eq:fdrag-final} scales proportional to the mass of the moving body
squared, and to the inverse of its velocity squared.  Furthermore, the dynamical
friction $F_\mathrm{DF}$ scales linearly with the gas mass density and is
proportional to the logarithm of the ratio of the spatial extents of the forming
wake and the shock's stand-off distance.

\subsection{The minimum spatial interaction scale}
The numerical experiments allow us to not only compute the total dynamical
friction acting on the body, but also to investigate the individual
contributions to the dynamical friction induced as a function of the distance to
the perturber.  The outcome of this analysis reveals that the region within the
stand-off distance $R_\mathrm{SO}$ of the shock does not contribute to the
dynamical friction of a gaseous medium (see Fig.~\ref{fig:radial-drag} and
associated main text for details).  At radii smaller than the stand-off
distance, a spherically symmetric stratified atmosphere forms around the moving
body, which due to its symmetry induces no net gravitational pull or
pressure force.  Only at radii above the stand-off distance, a wake of higher
density forms behind the moving body and induces a gravitational drag force.

Hence, in the well-known formulae by \citet{1971ApJ...165....1R} or
\citet{1999ApJ...513..252O}, it is the stand-off distance $R_\mathrm{SO}$ of the
shock, which determines the minimum length scale $s_\mathrm{min}$ required in
the so-called Coulomb-Logarithm $C_A = \ln(s_\mathrm{max}/s_\mathrm{min})$:
\begin{equation}
    s_\mathrm{min} = R_\mathrm{SO}~.
\end{equation}
The formation of a spherically symmetric atmosphere (with radius
$R_\mathrm{SO}$) around a gravitating object implies that the hydrodynamical
drag force vanishes in contrast to the non-gravitating case. But for very high
Mach numbers the shock moves very close to the object so that $R_\mathrm{SO}
\approx R$. In this case no spherically symmetric envelope around the object
formes and the hydrodynamical drag becomes important again.

\subsection{Relating the stand-off distance to the accretion radius}
In a second step, to allow an easy computation of the stand-off distance for
general dynamical friction problems, we derived the relation of the shock's
stand-off distance to the common definition of the accretion radius:
\begin{equation}
    R_\mathrm{A} = \frac{2 G M}{V_\infty^2}~,
\end{equation}
depending only on the mass $M$ and velocity $V_\infty$ of the perturber.  The
stand-off distance is directly proportional to the accretion radius and the
proportionality constant depends on the Mach number and the adiabatic index of
the gas only.
\begin{equation}
    R_\mathrm{SO} = \frac{g(\gamma)}{2} \, \frac{\mathcal{M}^2}{\mathcal{M}^2 - 1}~R_\mathrm{A} ~.
\end{equation}
Additionally, the geometrical radius $R$ of the moving body yields a lower limit
for the stand-off distance $(R_\mathrm{SO} \ge R)$.

Via an analytical estimate of the stand-off distance and further fitting of the
remaining free parameters, we gave an approximate relation of the dependence of
the shock's stand-off distance on the adiabatic index $\gamma$ of the gaseous
medium:
\begin{equation}
    g(\gamma) = 
    \frac{15}{8} \left\{
    \frac{(\gamma' + 1)^2}{4\gamma'} 
    \left[1 - \left(\frac{(\gamma'-1)}{\gamma'+1}\right)^{\gamma'-1}\right]^{-1} 
    - 1 - b \right\}
\end{equation}
with $\gamma' = \gamma + a$ and values of $a=0.1$ and $b=0.18$.

\subsection{An updated formula of dynamical friction}
Finally, we combined the derived scaling relations and the finding on the
stand-off distance to give an update of the well-known formula of dynamical
friction acting on a body of mass $M$ moving at supersonic speed $V_\infty$
through a gaseous homogeneous medium of mass density $\rho_\infty$:
\begin{equation}
F_\mathrm{DF} = 4\pi\rho_\infty \left(\frac{G M}{V_\infty} \right)^2 ~ \ln \left( \frac{s_\mathrm{max}}{R_\mathrm{SO}} \right)~,
\end{equation}
where $s_\mathrm{max}$ denotes the extent of the wake and $R_\mathrm{SO}$ can be
obtained using the derived formulae above.

\begin{acknowledgements}
We thank Chris Ormel for very helpful discussions.  R.~K.~acknowledges financial
support by the Emmy-Noether-Program of the German Research Foundation (DFG)
under grant no.~KU~2849/3-1.  Part of the numerical simulations were performed
on the bwGRiD cluster in T\"ubingen, which is funded by the Ministry for
Education and Research of Germany and the Ministry for Science, Research and
Arts of the state Baden-W\"urttemberg, and the cluster of the Forschergruppe FOR
759 funded by the DFG. All plots in this paper have been made with
the Python library matplotlib \citep{Hunter:2007}.
\end{acknowledgements}

\bibliography{Papers_DT,Papers_RK}{}
\bibliographystyle{aa}

\appendix

\section{Hydrostatic equilibrium stratification}
\label{app:strati}
\begin{figure}[htbp]
    \begin{center}
    \includegraphics[width=0.4\textwidth]{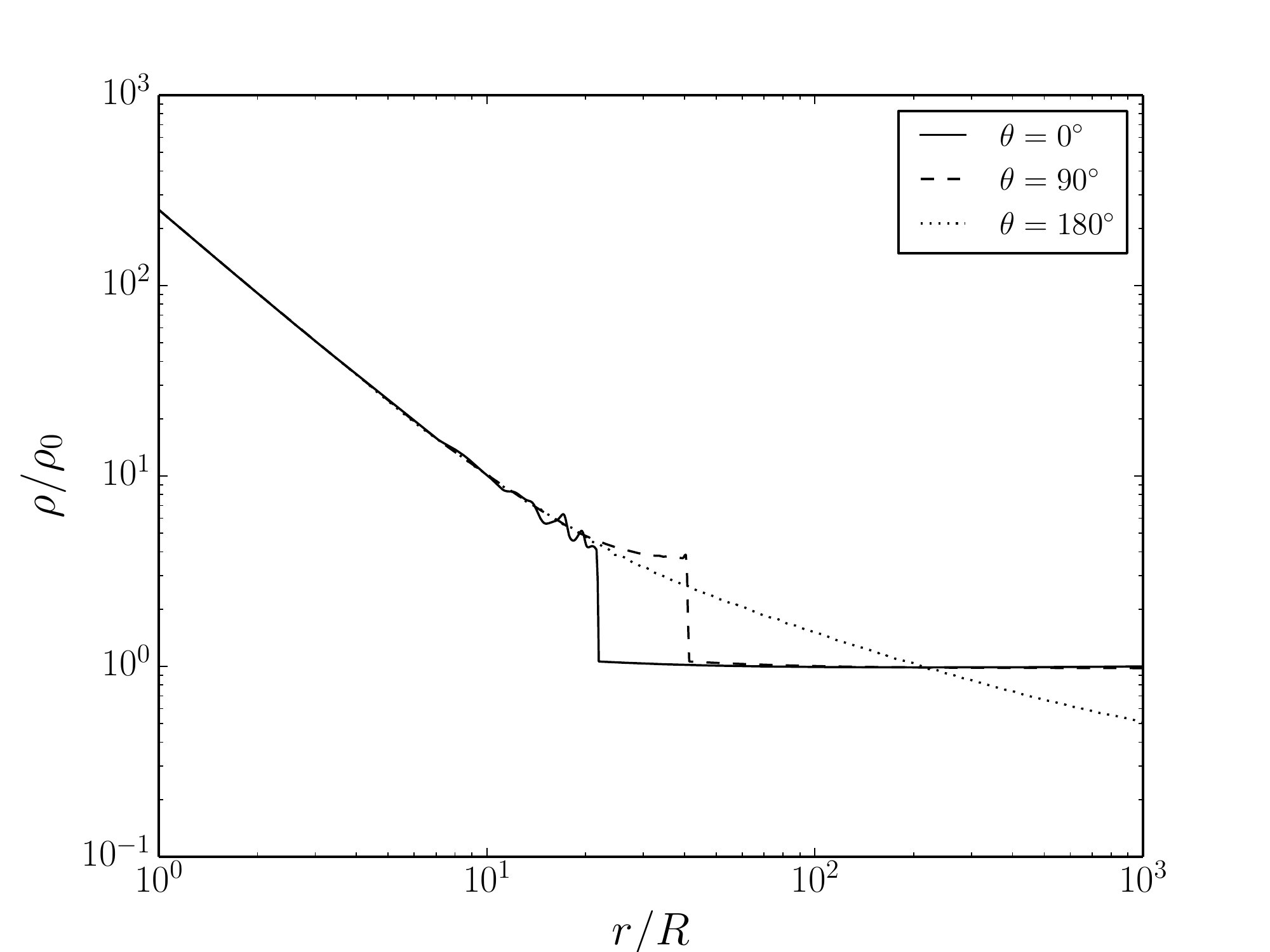}\\ 
    \includegraphics[width=0.4\textwidth]{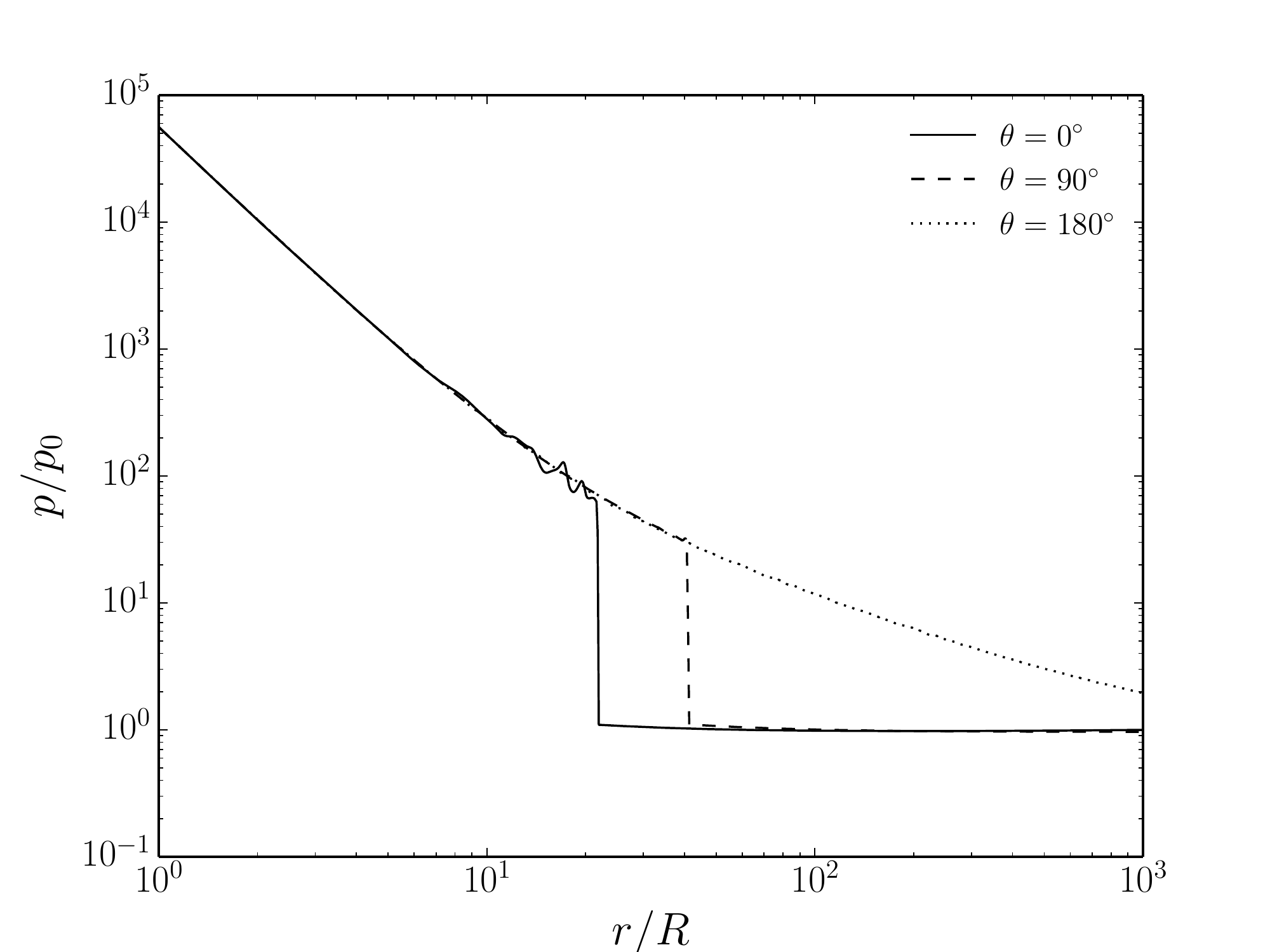}\\ 
    \includegraphics[width=0.4\textwidth]{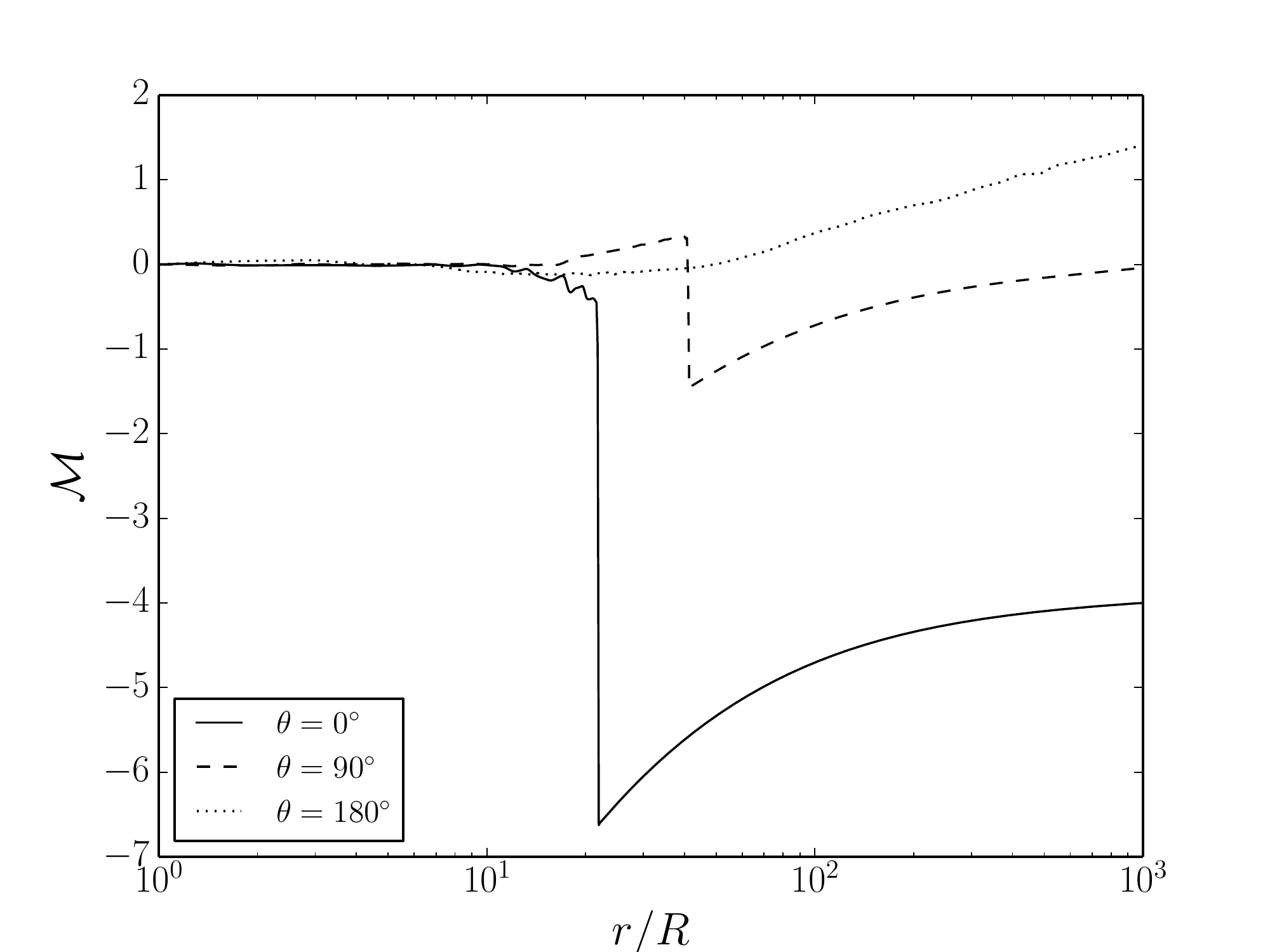}\\ 
    \includegraphics[width=0.4\textwidth]{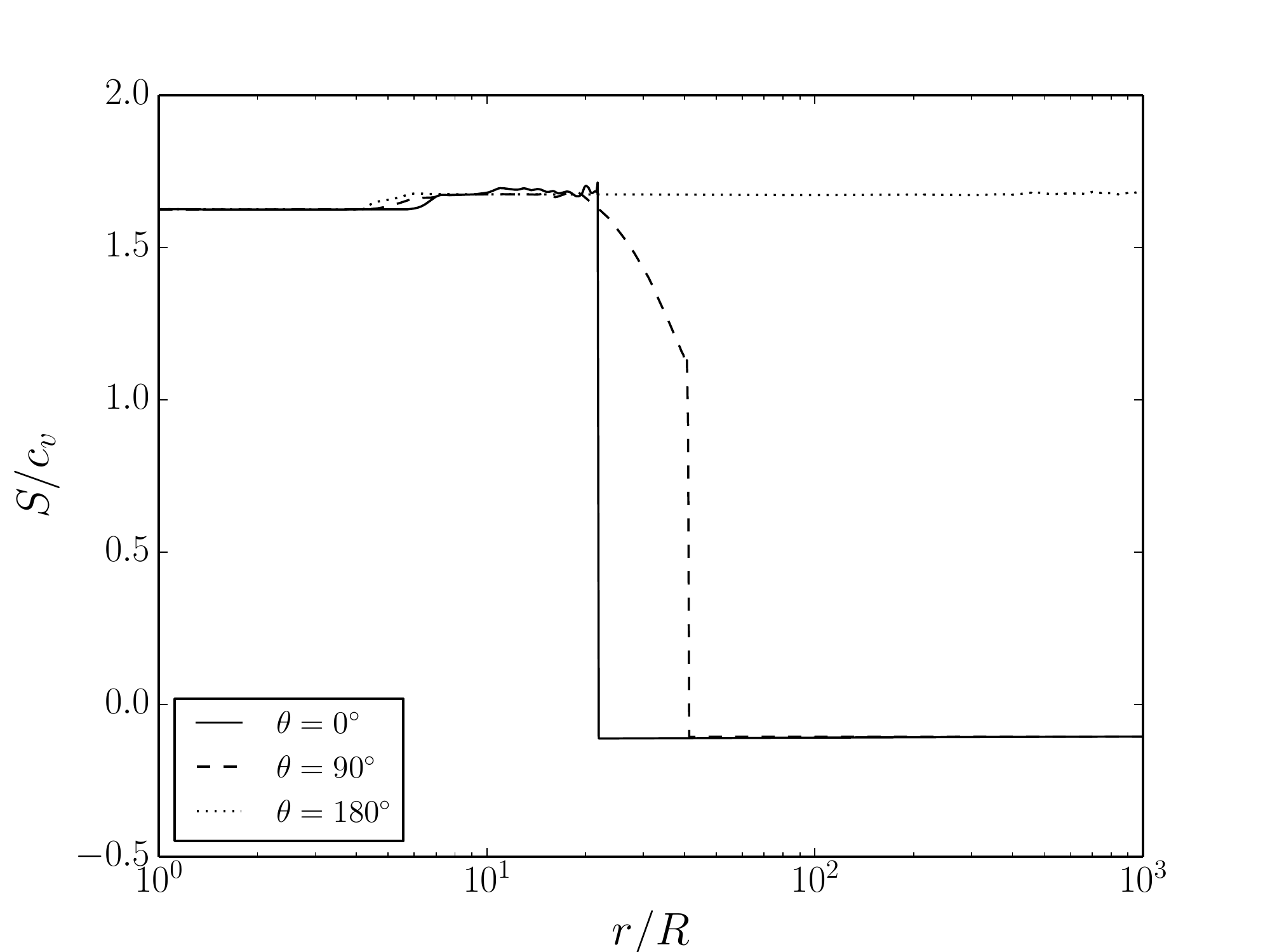}\\ 
    \end{center}
    \caption{Stratification around the moving object.  Values in front of the
        moving object ($\theta = 0$), perpendicular to its velocity ($\theta =
        90$), and behind the object ($\theta = 180$) are displayed.  From top to
        bottom, the panels show gas density, pressure, Mach number, and entropy
        of the flow.}
    \label{fig:cut_ref}
\end{figure}

As mentioned in the text, behind the shock front and around the object the
material is very subsonic and can be approximated by assuming a hydrostatic
equilibrium.  The morphology is illustrated in more detail in
Fig.~\ref{fig:cut_ref} where we show the stratification of different physical
quantities in different directions.  The data shown is extracted from a
simulation with the fiducial parameters as specified in Table~\ref{tab:sims}.

\begin{figure}[htbp]
    \begin{center}
    \includegraphics[width=0.48\textwidth]{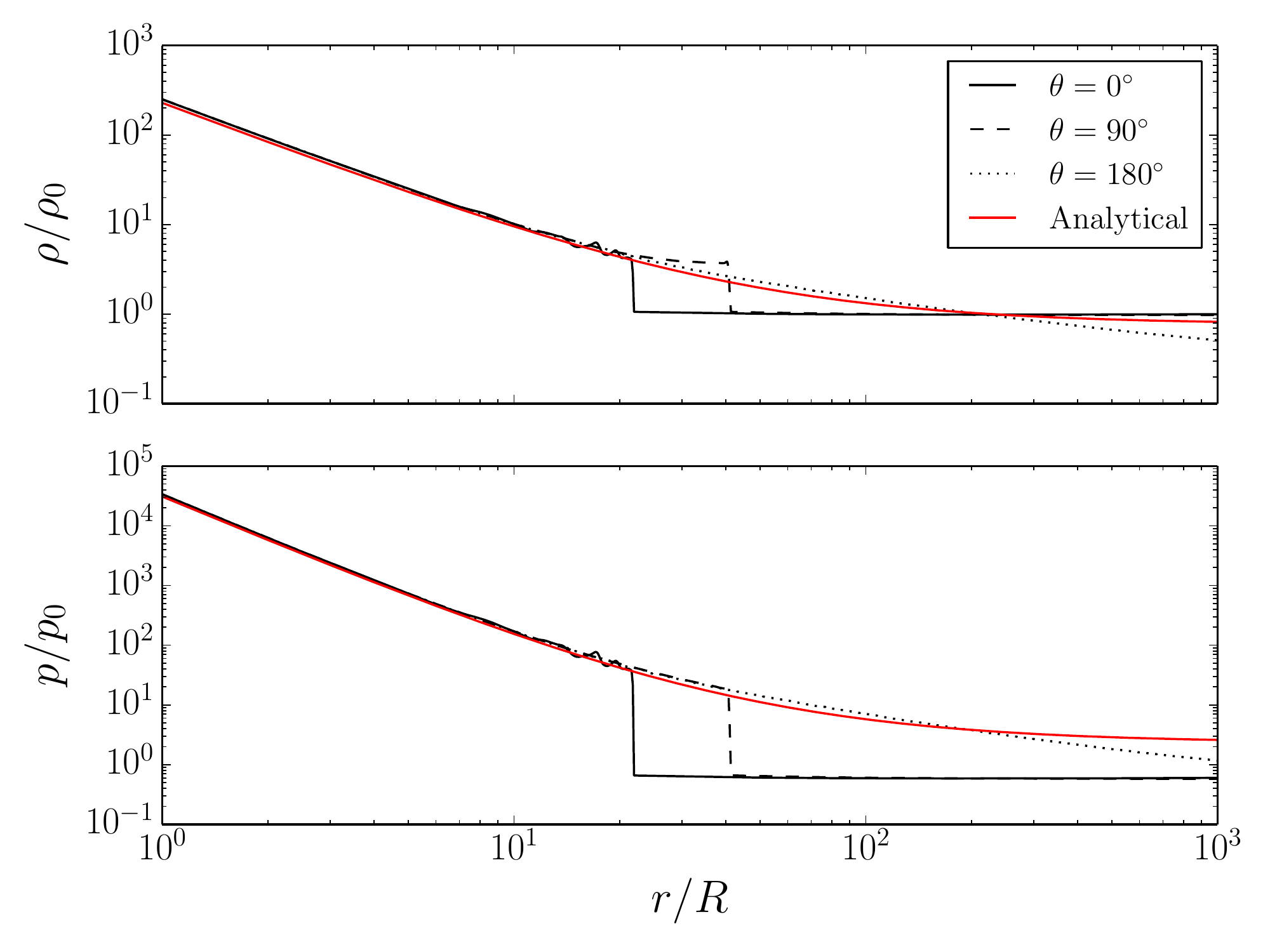}
    \end{center}
    \caption{Stratification of gas density (upper panel) and pressure (bottom
        panel) around the moving object.  As in the previous figure, values in
        front of the moving object ($\theta = 0$), perpendicular to its velocity
        ($\theta = 90$), and behind the object ($\theta = 180$) are displayed.
        The additional red lines denote the analytical estimates obtained from
        Eqs.~\eqref{eq:rhopost} and \eqref{eq:press}, respectively.  In the
        formula of the estimate, a value for the stand-off distance of
        $x_\mathrm{SO} = 0.533$ was used, in accordance with the final
        quasi-stationary state of the numerical experiment.}
    \label{fig:stratification}
\end{figure}
As clearly visible, within the stand-off radius, as given by the discontinuity
of the solid curve, all three cuts agree with each other, and hence the
stratification is spherically symmetric.  Because of the Mach number being very
close to zero it is hydrostatic as well.  Here, we derive analytical relations
for this stratification that will be used in the following section of the
appendix to obtain an analytical estimate on the shock's stand-off distance.

For spherical symmetry the equation of hydrostatic equilibrium reads
\begin{equation}\label{eq:hydrostat}
    \frac{1}{\rho} \doverdr{p} = - \doverdr{\phi}
\end{equation}
where $\phi$ is the gravitational potential. 
Now we use an adiabatic approximation for the pressure
\begin{equation}\label{eq:press}
    p = K \rho^\gamma\, ,
\end{equation}
where $K$ is a constant, and integrate from a reference radius $r_0$ to $r$,
yielding:
\begin{equation}\label{eq:rhor}
     \left( \frac{\rho}{\rho_0} \right)^{\gamma-1} = 
        1 - \frac{\gamma -1}{c_{0}^2} \, \left(  \phi(r)  - \phi_0 \right).
\end{equation}
Here the subscript $0$ refers to the reference radius $r_0$ where the values are
known.

As a next step, we want to analyze the bow-shock in front of the object.  We can
find the physical properties behind the shock from the standard jump conditions
\begin{align}
    \frac{\rho_2}{\rho_1} &= \frac{(\gamma+1)\Mach_1^2}{(\gamma-1)\Mach_1^2 + 2}
    \\
    \frac{p_2}{p_1} &= \frac{(\gamma+1)+2\gamma(\Mach_1^2 - 1)}{\gamma+1}
\end{align}
Here, the index '1' refers to the pre-shock values (supersonic regime) and the
index '2' to the post-shock (subsonic) regime.  These are valid in a reference
frame moving with the shock.  In our case the shock is stationary in the
co-moving frame of the body, and we can directly apply the jump conditions,
knowing the pre-conditions.  The simulations show that for the density,
pressure, and temperature the pre-shock values, at the the stand-off radius
$R_\mathrm{SO}$, are just the prescribed inflow values (at $\infty$), compare
also Fig.~\ref{fig:cut_ref}.

To obtain the pre-shock Mach-number $\Mach_1$ at this radius we use the
free-fall condition.  From the conservation of energy (kinetic and
gravitational), we obtain
\begin{equation}\label{eq:M1}
    \Mach_1 = \Mach_\infty  \, \left( 1 + \frac{1}{x_{SO}} \right)^{1/2}
\end{equation}
where the coordinate $x$ denotes the radial distance to the shock front in units
of the accretion radius, specifically $x_\mathrm{SO} =
R_\mathrm{SO}/R_\mathrm{A}$.  Eq.~\eqref{eq:M1} matches the simulation outcome
quite well, cp.~Fig.~\ref{fig:cut_ref}.

To obtain the (hydrostatic) post-shock density stratification around the object
we first use these normalizations to obtain the general relation
\begin{equation} \label{eq:rhopost}
    \rho(x) = \rho_0 \left[1 - \frac{\gamma -1}{c_{0}^2/c_\infty^2} \frac{\mathcal{M}_\infty^2}{2} \left( \frac{1}{x_0} -\frac{1}{x} \right) \right]^{1/\gamma-1}
\end{equation}
with an arbitrary point $x_0$ where $\rho_0$ and $c_{0}$ are known values at the
reference radius.  If we choose now $x_0 = x_\mathrm{SO}$, the other quantities
correspond to the desired postshock quantities and can be obtained from the jump
conditions, using $\Mach_1$ from Eq.~\eqref{eq:M1}.  The required shock-position
$x_\mathrm{SO}$ enters at this point as a free parameter, obtained from the
numerical solution.
 
\section{An estimate for the stand-off distance} \label{app:stand-off}
Because the stratification around the object is nearly hydrostatic, one can
obtain some limits on $x_\mathrm{SO}$.  Here, we use the strong shock
conditions, i.e. $\Mach_1 \rightarrow \infty$ and obtain for the jumps in
density and pressure
\begin{align}
    f_\rho \equiv  \frac{\rho_2}{\rho_1} &= \frac{\gamma+1}{\gamma-1} \\
    f_p  \equiv  \frac{p_2}{p_1} &= \frac{2 \gamma \Mach_1^2}{\gamma+1}
\end{align}
and hence for the post-shock soundspeed (in units of $c_\infty$)
\begin{equation} \label{eq:app-cs}
    c_\mathrm{SO}^2 = \frac{f_p}{f_\rho} 
                    = \frac{2\gamma \Mach_1^2}{(\gamma+1)^2} \, (\gamma -1) \,.
\end{equation}
Now, we assume that for large distances, $x \rightarrow \infty$, the density of
the post-shock stratification lies below the unperturbed density $\rho (x
\rightarrow \infty) \leq \rho_\infty$ as seen in the simulations.  This fact is
displayed in Fig.~\ref{fig:stratification} where the analytical post-shock
stratification (red solid curve) lies for large radii just below the unperturbed
value $\rho_\infty$.  Inserting Eq.~\eqref{eq:app-cs} in Eq.~\eqref{eq:rhopost}
yields
\begin{equation}
    1 \geq \frac{\gamma+1}{\gamma-1}  \, \left[  1  - \frac{(\gamma+1)^2
        \Mach_\infty^2}{2 \gamma \Mach_1^2 2}
        \frac{1}{x_\mathrm{SO}} \right]^{1/\gamma-1}
\end{equation}
with $\Mach_1$ from Eq.~(\ref{eq:M1}). 
Solving for $x_\mathrm{SO}$ finally leads to
\begin{equation} \label{eq:xso_gamma}
    x_\mathrm{so} \leq \frac{(\gamma+1)^2}{4 \gamma} \,
      \left[1 - \left(\frac{\gamma-1}{\gamma+1} \right)^{\gamma-1} 
      \right]^{-1}  - 1 \,.
\end{equation}
This function is shown in Fig.~\ref{fig:gam-rso} in comparison to data from
numerical simulations.
While the relation \eqref{eq:xso_gamma} cannot give the exact value for
$R_\mathrm{SO}$ it nevertheless provides a reasonable upper limit for the
stand-off distance, and follows the trend of the obtained numerical results.

\section{Convergence study}
\label{sect:res_test}
We ran simulations using varying grid sizes from roughly $10^3$ up to $10^6$
grid cells in total (more precisely, we varied the number of grid cells from
$N_\mathrm{r} \times N_\mathrm{\theta} = 50 \times 23 = 1150$ up to
$N_\mathrm{r} \times N_\mathrm{\theta} = 1720 \times 790 = 1.3588  \times 10^6$
grid cells).  The inner radial boundary of these simulations was chosen to $R =
0.01 \mbox{ R}_\odot$.  Other simulation parameters correspond to the ``fiducial
setup'' given in Table~\ref{tab:sims}.  From the final quasi-stationary state,
we determine the two most relevant physical quantities of our investigation,
namely the shock's stand-off distance $R_\mathrm{SO}$ and the dynamical friction
$F^\mathrm{numerical}_\mathrm{DF}$ acting on the moving body.

\begin{figure}[htbp]
    \begin{center}
    \includegraphics[width=0.49\textwidth]{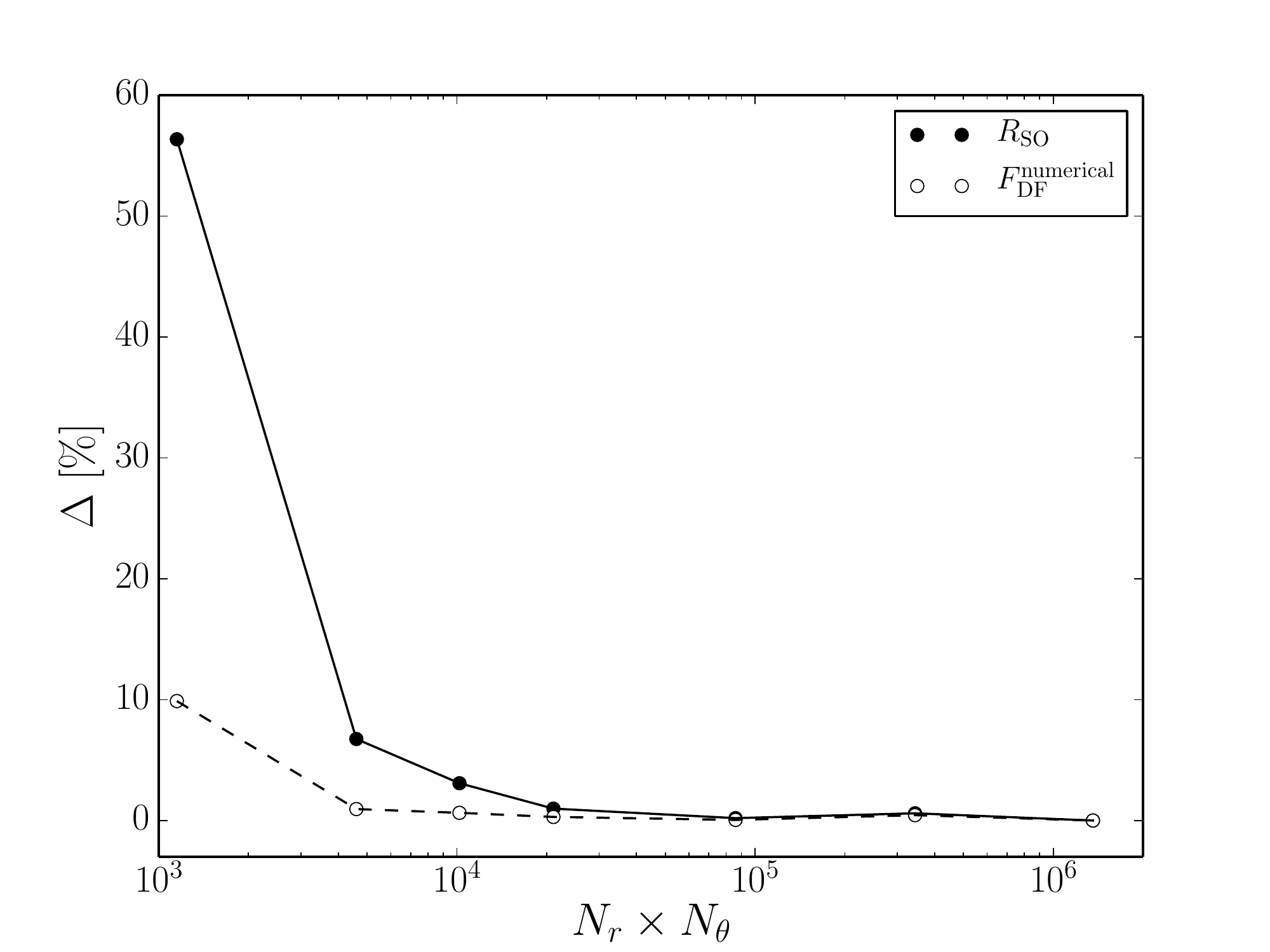}
    \end{center}
    \caption{Convergence test of the numerical setup.  Values on the horizontal
        axis denote the total number of grid cells $N_\mathrm{r} \times
        N_\mathrm{\theta}$.  The vertical axis shows the relative deviation of
        the resulting stand-off distance $R_\mathrm{SO}$ (solid circles) and the
        dynamical friction $F^\mathrm{numerical}_\mathrm{DF}$ (dashed circles) from its
        corresponding value of the highest resolution run.}
    \label{fig:res_test}
\end{figure}

Results of this convergence study are presented in Fig.~\ref{fig:res_test}.
Deviations are given as relative differences to the values from the highest
resolution simulation using more than $10^6$ grid cells.  The numerical results
show a clear convergence trend.  For numerical grids larger than $10^5$ grid
cells, the deviations become negligibly small.  To be on the very safe side, we
chose $N_\mathrm{r} \times N_\mathrm{\theta} = 860 \times 400 = 3.44 \times
10^5$ grid cells as our default resolution for the simulations performed.
Simulations with a larger inner radial boundary of $R = 0.1 \mbox{ R}_\odot$
require less grid cells in the radial direction, especially due to the
logarithmic grid in the radial direction.  Nonetheless, we also use here a
comparable grid size of $N_\mathrm{r} \times N_\mathrm{\theta} = 700 \times 480
= 3.36 \times 10^5$ grid cells as our default resolution.

\end{document}